\documentclass[prl,aps,twocolumn,superscriptaddress,longbibliography]{revtex4-2}

\usepackage[utf8x]{inputenc}
\usepackage[T1]{fontenc}
\usepackage{amssymb}
\usepackage{amsmath}
\usepackage{graphicx}
\usepackage{bbm}
\usepackage{psfrag}

\usepackage{latexsym}
\usepackage{color}
\usepackage[dvipsnames]{xcolor}
\usepackage{hyperref}
\hypersetup{colorlinks=true, citecolor=blue, urlcolor=blue, linkcolor=blue,urlcolor=cyan}
\usepackage{tensor}
\usepackage{amsfonts}
\usepackage{amsmath}
\allowdisplaybreaks[4]        
\usepackage{amssymb}
\usepackage{euscript}           
\usepackage{tensor}        
\usepackage{amsthm} 
\usepackage{graphicx}
\usepackage{subfigure}
\usepackage{bbm}
\usepackage[header,title,page,titletoc]{appendix}  
\usepackage[most]{tcolorbox}
\tcbuselibrary{skins, breakable, theorems}
\usepackage{verbatim}

\newcommand{\eg}{{\it e.g.,}\ }
\newcommand{\ie}{{\it i.e.,}\ }
\newcommand{\viz}{{\it viz,}\ }
\newcommand{\mt}[1]{\textrm{\tiny #1}}

\renewcommand{\[}{\left[}

\newcommand{\mC}{\mathcal{C}}

\newcommand{\GN}{G_\mt{N}}

\newcommand{\THH}{T_\mt{H}}

\newcommand{\mD}{\mathcal{D}}

\newcommand{\talpha}{\tilde{\alpha}}

\begin{document}
\preprint{RIKEN-iTHEMS-Report-25}

\author{Masamichi Miyaji}\email{masamichi.miyaji@gmail.com}

\affiliation{\it RIKEN Center for Interdisciplinary Theoretical and Mathematical Sciences (iTHEMS),
RIKEN, 2-1 Hirosawa, Wako, Saitama, 351-0198, Japan}

\author{Shan-Ming Ruan}\email{ruanshanming@pku.edu.cn}
\affiliation{\it School of Physics and Center of High Energy Physics, Peking University, Beijing 100871, China}
\author{Shono Shibuya}\email{shibuya.shono.n8@s.mail.nagoya-u.ac.jp}
\author{Kazuyoshi Yano}\email{yano.kazuyoshi.h8@s.mail.nagoya-u.ac.jp}
\affiliation{Department of Physics, Nagoya University, Nagoya, Aichi 464-8602, Japan}

\title{Universal Time Evolution of Holographic and Quantum Complexity}

\begin{abstract}
Holographic complexity, as the bulk dual of quantum complexity, encodes the geometric structure of black hole interiors. Motivated by the complexity=anything proposal, we introduce the spectral representation for generating functions associated with codimension-one and codimension-zero holographic complexity measures. These generating functions exhibit a universal slope-ramp-plateau structure, analogous to the spectral form factor in chaotic quantum systems. In such systems, quantum complexity evolves universally, displaying long-time linear growth followed by saturation at late times. By employing the generating function formalism, we show that this universal behavior has two origins: a particular pole structure of the matrix elements of the generating functions in the energy eigenbasis and random matrix universality in spectral statistics. Using the residue theorem, we prove that the existence of this pole structure is a necessary and sufficient condition for the linear growth of holographic complexity measures. Furthermore, we show that the late-time saturation plateau arises directly from the spectral level repulsion, a hallmark of quantum chaos.
\end{abstract}

\maketitle

\noindent \emph{1.~\textbf{Introduction}.—}
Quantum gravity is confronted by profound tensions between the smooth geometry of classical spacetime and the discrete and finite nature of quantum Hilbert space. 
A prominent illustration of this tension appears in the spectral statistics of black holes. One notable example is Maldacena’s black hole information paradox \cite{Maldacena:2001kr}, 
which can be characterized by the time evolution of holographic correlators, such as two-sided correlation functions $\langle \mathrm{TFD}_\beta|  \mathcal{O}(t_{\mt{L}}) \mathcal{O}(t_{\mt{R}}) | \mathrm{TFD}_\beta \rangle $. Its spectral representation is given by
\begin{equation}\label{eq:twopointfuncion}
\begin{split}
\frac{1}{Z} \sum_{i, j} e^{-\frac{\beta}{2}\left(E_i+E_j\right)} e^{-i T\left(E_i-E_j\right)}\left|\left\langle \mathbf{E}_j\right| \mathcal{O}\right| \mathbf{E}_i\rangle|^2 \,,
\end{split}
\end{equation}
with the time shift defined as $T = t_{\mt{L}} + t_{\mt{R}}$ \footnote{The time coordinates $t_{\mt{L}}$ and $t_{\mt{R}}$ on the left and right boundaries both increase upwards (or downwards).} and $Z = \mathrm{Tr} \, e^{-\beta H}$ representing the partition function. While semiclassical gravity predicts an indefinite decay of these correlation functions, a unitary and finite quantum theory requires erratic fluctuations and a late-time plateau. This apparent contradiction is succinctly captured by the spectral form factor (SFF), which exhibits a characteristic slope-ramp-plateau structure \footnote{The characteristic timescale marking the onset of the linear ramp is the Thouless time $T_{\rm Th}$ \cite{Sonner:2017hxc,Gharibyan:2018jrp,Altland:2020ccq,Altland:2022xqx,Orman:2024mpw}.}\nocite{Sonner:2017hxc,Gharibyan:2018jrp,Altland:2020ccq,Altland:2022xqx,Orman:2024mpw}
in chaotic quantum systems, \eg the Sachdev–Ye–Kitaev (SYK) model \cite{Sachdev_1993,KitaevTalks,Cotler:2016fpe}  and Jackiw-Teitelboim (JT) gravity \cite{Saad:2018bqo,Saad:2019lba, Saad:2019pqd,Okuyama:2020ncd,Blommaert:2022lbh,Saad:2022kfe,Okuyama:2023pio,Griguolo:2023jyy}. 

\begin{figure}[t]
\centering\includegraphics[width=3.37in]{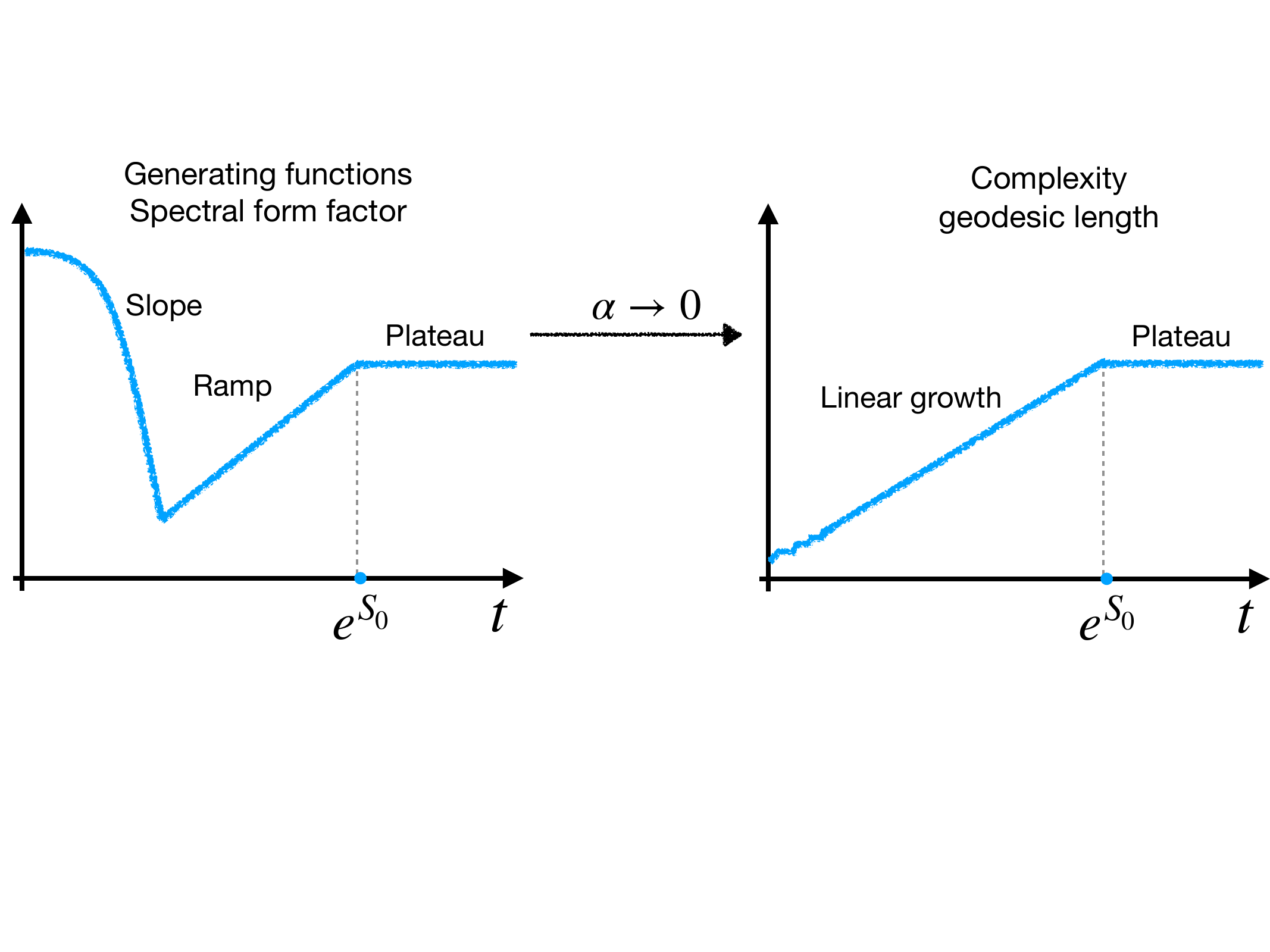}
	  \caption{Left: Typical time evolution of generating functions of complexity $\langle e^{-\alpha \mC} \rangle$ presents the well-known slope-ramp-plateau structure. Right: The time evolution of complexity measures could be derived from its generating function in the limit $\alpha \to 0$.}\label{fig:GeneratingFunctions}
\end{figure}

\begin{figure}[h]
	\centering\includegraphics[width=3.3in]{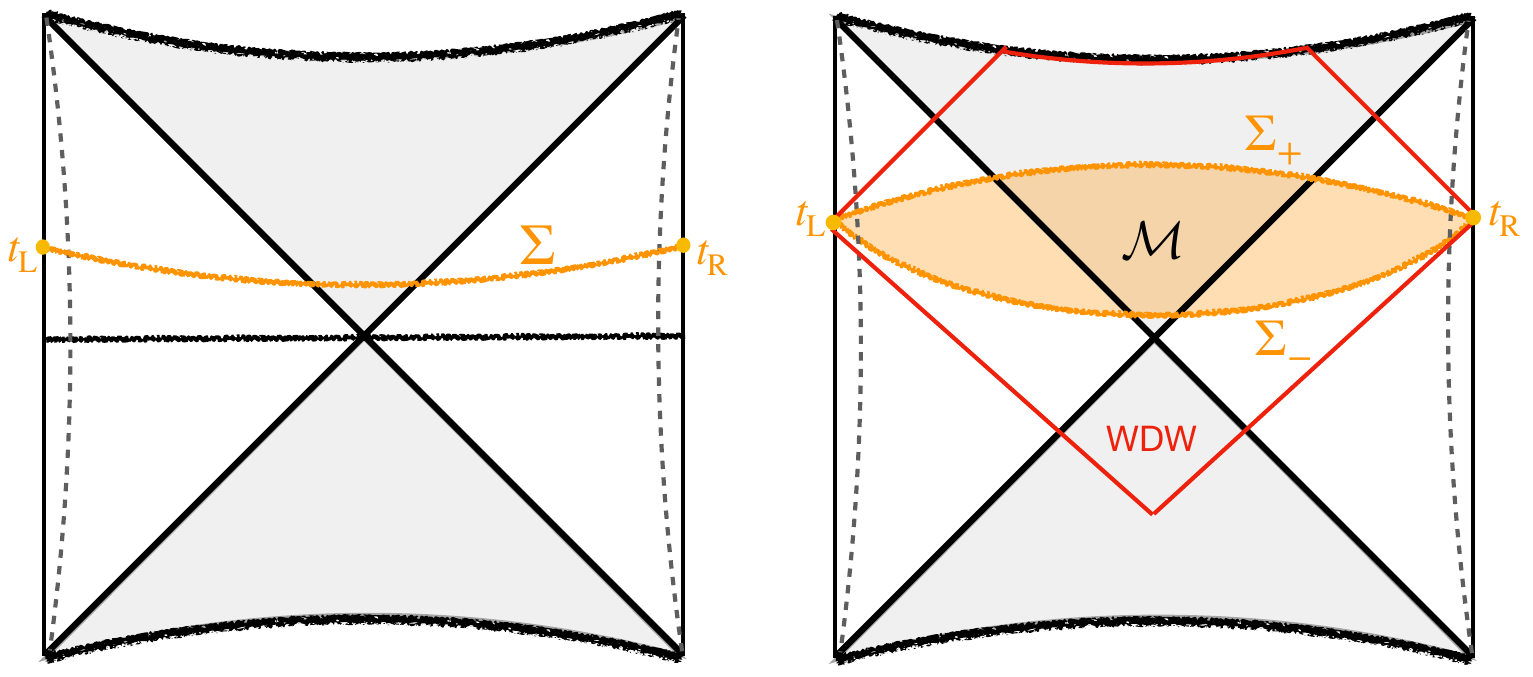}
	\caption{Complexity=anything proposal. Left: Codimension-one observables are supported on extremal hypersurface $\Sigma$. Right: Codimension-zero observables are defined by extremal subregion $\mathcal{M}$ with boundaries $\Sigma_\pm$.}\label{fig:CAnything}
\end{figure}

Another striking manifestation of this tension is Susskind's wormhole size paradox. Specifically, the wormhole's size in classical spacetime, \eg the volume of the black hole interior or Einstein-Rosen bridge (see Fig.~\ref{fig:CAnything}), grows linearly without bound at late times. According to the complexity=volume conjecture ~\cite{Susskind:2014rva,Stanford:2014jda, Brown:2015bva,Brown:2015lvg,Caputa:2017urj,Caputa:2017yrh, Brown:2019rox,Belin:2021bga,Belin:2022xmt}
or the related proposal identifying volume with the information metric \cite{Miyaji:2015woj,Miyaji:2016fse,Belin:2018bpg}, this infinite growth appears to contradict the finiteness of the Hilbert space dimension in quantum gravity. Brown and Susskind conjectured \cite{Brown:2017jil} that quantum complexity \footnote{We refer to the measures of complexity for boundary field theory as quantum complexity.} in chaotic quantum systems exhibits a long-time linear growth \cite{Haferkamp:2021uxo} and eventually transitions to a plateau after a timescale $e^{S_0}$, as depicted in Fig.~\ref{fig:GeneratingFunctions}. This transition was first holographically realized in JT gravity by incorporating non-perturbative quantum corrections to geodesic lengths \cite{Iliesiu:2021ari}, see also \cite{Alishahiha:2022nhe,Iliesiu:2024cnh,Blommaert:2024ftn, Miyaji:2024ity,Balasubramanian:2024lqk,geodesic, Sato:2025uli,Miyaji:2025ucp,Gautason:2025ryg}. For more recent progress on complexity measures, see \cite{Baiguera:2025dkc} for a review.

Remarkably, these two paradoxes are fundamentally connected via the holographic dictionary. In particular, the geodesic approximation of holographic correlation functions \cite{Balasubramanian:1999zv,Louko:2000tp} implies that late-time saturation in both wormhole size and quantum correlators arises from the same underlying mechanism, governed by quantum chaos and random matrix universality. This connection was explicitly established in~\cite{geodesic} via generating functions for geodesics in JT gravity, as summarized in Fig.~\ref{fig:GeneratingFunctions}.

In this Letter, we provide a universal resolution to the wormhole size paradox by introducing generating functions $\langle e^{-\alpha \mC}\rangle$ for holographic complexity measures $\mC$. In light of the complexity=anything proposal \cite{Belin:2021bga,Belin:2022xmt,Jorstad:2023kmq} which suggests that there exist infinitely many analogous holographic complexity measures (see Fig.~\ref{fig:CAnything}), our framework applies universally to both codimension-zero and codimension-one gravitational observables within a generic holographic gravity setting. By employing spectral representations of generating functions, we explicitly identify their universal slope-ramp-plateau structure \footnote{In contrast to the spectral form factor \cite{prange1997spectral}, the generating functions of complexity $\langle e^{-\alpha \mC}\rangle $ defined in Eq.~\eqref{eq:complete} are expected to be self-averaging in the large-entropy limit.}\nocite{prange1997spectral}. Moreover, we prove that the universal time evolution of quantum or holographic complexity is governed by spectral correlations dictated by random matrix universality, as well as a specific pole structure in the matrix elements of these generating functions expressed in the energy eigenbasis.

In holographic quantum gravity, we generally assume that the Hilbert space of black hole microstates $\mathcal{H}_{\mt{BH}}$ is finite and discrete with a small but finite $\GN$. However, as a powerful approximation, we expand the Hamiltonian eigenbasis using a continuous spectrum $E_i$ and generalized eigenstates $|E_i \rangle$ \footnote{We adopt the convention for energy eigenstates as $\langle E_i|E_j \rangle \equiv \frac{\delta(E_{ij})}{e^{S_0} \bar{D}(E_i)}$.}, which are rigorously defined in the associated \emph{rigged Hilbert space}. In this distributional sense, they allow for a resolution of the identity operator:
\begin{equation}\label{eq:Identity}
 \hat{\mathbbm{1}} = e^{S_0} \int_{E_0 - \Delta E/2}^{E_0 + \Delta E/2} dE \, D(E) \, |E\rangle \langle E|  \,, 
\end{equation}  
where we focus on a microcanonical ensemble characterized by a finite energy window centered around $E_0$, \ie $E_i \in \left[E_0 - \frac{\Delta E}{2}, E_0 + \frac{\Delta E}{2}\right]$. Here, $D(E)$ denotes the rescaled density of states, and $S_0$ is the coarse-grained entropy characterizing the total dimension of the Hilbert space. Of particular interest to holography is the microcanonical thermofield double (TFD) state,
\begin{equation}\label{eq:TFD}
\begin{split}
    |\text{TFD}(t)\rangle
    &=\frac{e^{S_0}}{\sqrt{Z}}\int_{E_0-\Delta E/2}^{E_0+\Delta E/2}dE~  D(E) \,  e^{-iEt} \, |E\rangle \,.
\end{split}
\end{equation}
which is dual to a two-sided AdS black hole \footnote{From the viewpoint of the dual boundary theory, the energy eigenstates $|E\rangle$ can be expressed as products of left- and right-sided states, \ie, $|E\rangle = |\mathbf{E}\rangle_{\mt{L}} \otimes |\mathbf{E}\rangle_{\mt{R}}$. In this paper, however, we restrict ourselves to the diagonal subspace spanned by $|E\rangle$ and do not explicitly utilize the product structure.}. The normalization factor $Z \approx e^{S_0}\bar{D}(E_0)\Delta E$ represents the total number of microcanonical states, with $\bar{D}(E_0)$ being the ensemble-averaged density of states. A fundamental timescale for finite quantum systems is the Heisenberg time $T_{\mt{H}}$, defined as the inverse of the mean energy level spacing, namely 
\begin{equation}
T_{\mt{H}} : =  \frac{2\pi \hbar}{\text{average energy spacing} }=2\pi e^{S_0} \bar{D}(E_0) \,. 
\end{equation}
We will find that this timescale marks the transition in the time evolution of complexity from a linear growth regime to a late-time plateau, as illustrated in Fig.~\ref{fig:GeneratingFunctions}.

\vspace{4pt}
\noindent \emph{2.~\textbf{Codimension-one observables and spectral decomposition}.—}
To explicitly illustrate the time evolution of holographic complexity, we first consider codimension-one holographic complexity measures. Such codimension-one gravitational observables, denoted by $\mC$ \cite{Belin:2021bga}, can be interpreted as generalized volume functionals \footnote{These generalized volume functionals naturally appear in the context of holographic complexity in higher-curvature theories of gravity, see \eg \cite{Bueno:2016gnv,Hernandez:2020nem}.}\nocite{Bueno:2016gnv,Hernandez:2020nem} defined on the extremal hypersurface $\Sigma(t_{\mt{L}},t_{\mt{R}})$ (see Fig.~\ref{fig:CAnything}). Analogous to the geodesic-length operator studied in \cite{geodesic}, we formally define the \emph{operator of gravitational observables} corresponding to holographic complexity measures by
\begin{equation}
 \hat{\mC} :=  \sum_{\mC-\text{basis}} \mC \, | \mC \rangle \langle \mC |= \int \mC | \mC \rangle \langle \mC | \, d\mC \,,
\end{equation}
where $| \mC \rangle$ represents the quantum state with a fixed generalized volume $\mC$. However, the holographic complexity operator $\hat{\mC}$ is not always well-defined, because the set of states ${| \mC \rangle}$ can become over-complete due to quantum corrections from Euclidean wormholes in the gravitational path integral. In such cases, the complexity operator develops divergences and can deviate drastically from classical expectations \cite{geodesic}. A natural resolution is to select a complete orthonormal basis (denoted by ${|\mC_{O}\rangle}$), which can be obtained systematically via the Gram–Schmidt orthogonalization procedure \cite{Miyaji:2024ity,Miyaji:2025ucp}. Correspondingly, one can define a complexity operator in terms of the resulting discrete basis,
\begin{equation}\label{eq:defineCO}
    \hat{\mC}_O:=\sum_{\text{orthonormal basis}} \mC \, |\mC_O\rangle\langle\mC_O| \,. 
\end{equation}
This construction generalizes the notion of Krylov complexity, which has been extensively studied in recent literature ~\cite{Caputa:2021sib,Balasubramanian:2022tpr,Rabinovici:2022beu,Lin:2022rbf,Bhattacharjee:2022ave,Balasubramanian:2022dnj,Alishahiha:2022anw,Erdmenger:2023wjg,Hashimoto:2023swv,Caputa:2023vyr,Rabinovici:2023yex,Nandy:2024htc,Camargo:2024deu,Nandy:2024zcd,Balasubramanian:2024lqk,Ambrosini:2024sre,Baiguera:2025dkc,Rabinovici:2025otw}. Details of this approach are presented in the End Matter.

To obtain physically meaningful expectation values of holographic complexity measures $\mC$, a complementary approach was introduced in \cite{Iliesiu:2021ari,geodesic}. Instead of defining the complexity operator directly, one may focus on the regulated operator
 \begin{equation}
    \widehat{e^{-\alpha \mC }} : = \int e^{-\alpha \mC} | \mC \rangle \langle \mC | \, d\mC   \,, 
 \end{equation}
which is well defined even when ${|\mC \rangle}$ is over-complete, since the parameter $\alpha$ effectively acts as a regulator. The corresponding expectation values, namely 
\begin{equation}
 \langle e^{-\alpha \mC}\rangle : = \langle \text{TFD}(t)|\,  \widehat{e^{-\alpha \mC }} \, |\text{TFD}(t)\rangle \,,
\end{equation}
serve as generating functions for holographic complexity measures $\mC$. We then subtract the unphysical divergence by imposing that the generating operator reduces to the identity in the limit $\alpha \to 0$, \ie
\begin{equation}\label{eq:complete}
  \lim_{\alpha \to 0}  \widehat{e^{-\alpha \mC }} =    \hat{\mathbbm{1}} =  \sum_{\text{complete basis}}  | \mC \rangle \langle \mC |   \,, 
\end{equation}
where the sum is taken over a complete $\mC$-basis.

The main result of this Letter is that holographic complexity measures $\mC$ and their generating functions exhibit a universal time evolution that is fixed by a small set of holographic inputs, without requiring detailed knowledge of the microscopic structure of the holographic system or of the precise definition of $\mC$. The central observation is that the generating function admits a spectral decomposition:  
\begin{equation}\label{eq:spectraltwo}
  \frac{e^{2S_0}}{Z} \int d\bar{E}dE_{ij} \, e^{-iE_{ij}t} \overline{\langle  D(E_i) D(E_j) \rangle}  \langle E_i | \widehat{e^{-\alpha \mC }}| E_j \rangle  \,,
\end{equation}
with $\bar{E}=\frac{E_i +E_j}{2}, E_{ij}=E_i -E_j$. This decomposition makes manifest that the time evolution of the generating function is governed by two independent ingredients: the matrix elements $\langle E_i | \widehat{e^{-\alpha \mC}} | E_j \rangle$ and the spectral correlator $\overline{\langle D(E_i) D(E_j)\rangle}$. The universal time evolution follows from the fact that, in holographic systems, these two ingredients are constrained by the following two holographic inputs:
\begin{itemize}
\item[(i)] \emph{Holographic complexity grows linearly in time in the semiclassical regime.}
\item[(ii)] \emph{Black holes are dual to quantum chaotic systems.}
\end{itemize}
The first requirement imposes a strong constraint on the analytic structure of the matrix elements $\langle E_i | \widehat{e^{-\alpha \mC}} | E_j \rangle$ of the generating function. As discussed in later sections, we find that linear growth occurs if and only if the matrix elements after analytic continuation develop a particular pole structure, \ie 
\begin{equation}\label{eq:pole}
\begin{split}
&\langle E_i | \widehat{e^{-\alpha \mC }}  | E_j \rangle \sim \left( \frac{\tilde{\alpha}}{e^{S_0} \pi \bar{D}(\bar{E})} \right)  \frac{1}{(\tilde{\alpha} + i E_{ij} )(\tilde{\alpha} - i E_{ij})} + \cdots
\end{split}
\end{equation}
Here, we focus on the regime with $\alpha\sim E_{ij}\ll 1$ and define $\tilde{\alpha}=M\alpha+\mathcal{O}(\alpha^2)$. The essential feature is the presence of a simple pole located precisely on the imaginary axis at $E_{ij}=\pm i\tilde{\alpha}$. The overall coefficient is fixed by the completeness of the $\mC$-basis (cf. Eq.~\eqref{eq:complete}) together with the normalization of the energy eigenstates.

The second input instead constrains the ensemble-averaged spectral correlator. One key universal feature of quantum chaotic systems is \emph{level repulsion} which expresses the fact that nearby energy levels avoid one another. Mathematically it implies that the off-diagonal part of the $n$-point spectral correlator vanishes as two levels coincide, \eg 
\begin{equation}\label{eq:offdiag}
\lim_{E_{ij}\to 0} \overline{\langle D(E_i) D(E_j) \rangle}_{\rm off\text{-}diag}=0 \,.
\end{equation}
As we will show, this property ensures that the linear growth dictated by the pole structure does not persist indefinitely, but instead crosses over to a late-time plateau.

\begin{figure}[t]
	\centering	\includegraphics[width=2in]{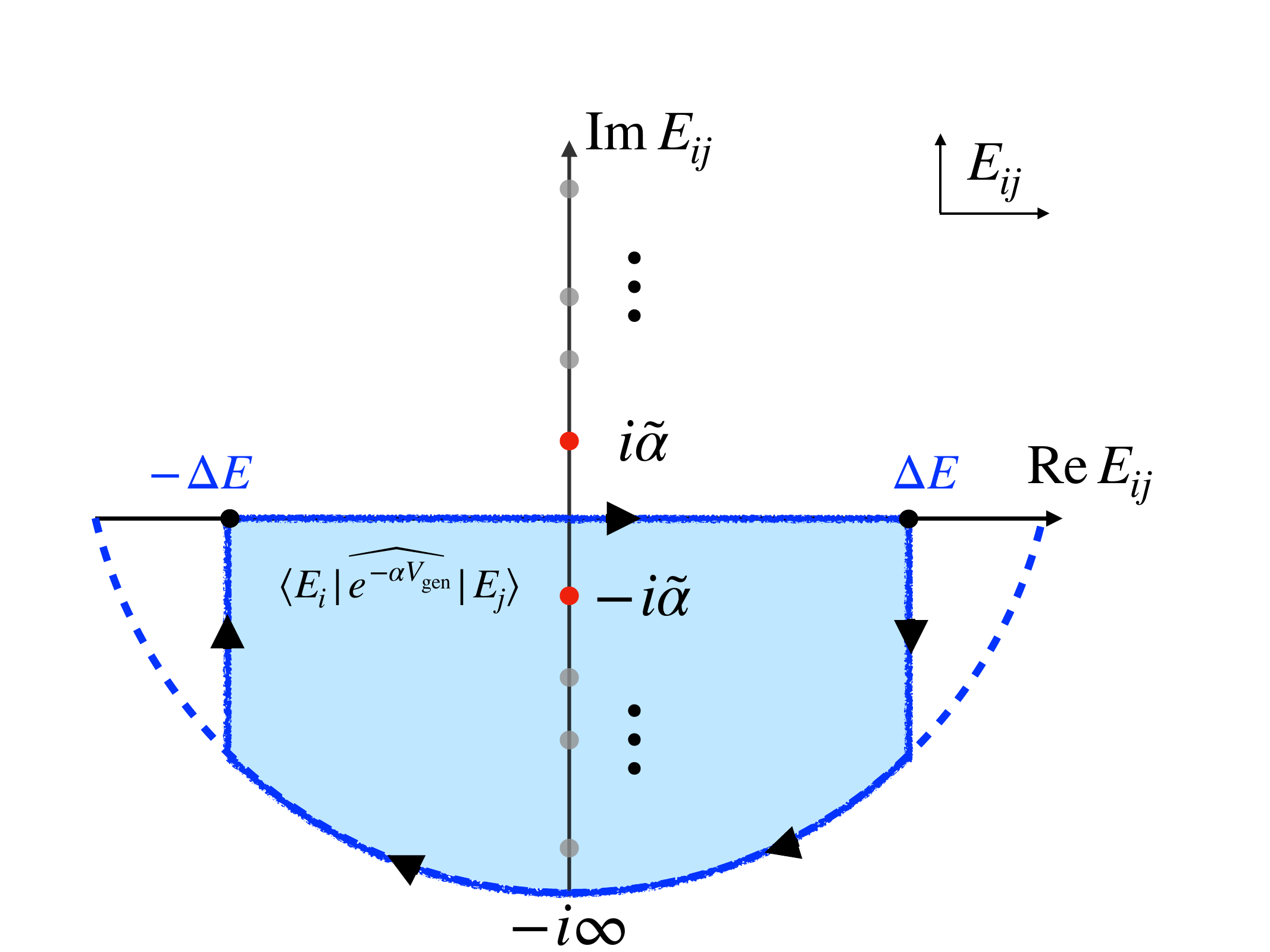}
	\caption{The integration contour corresponding to both the classical contributions and quantum corrections at late times.}\label{fig:pole01}
\end{figure}

\vspace{4pt}
\noindent \emph{3.~\textbf{Explicit example from GUE}.—}To make our results more concrete, we now illustrate the universal time evolution by using an explicit example. It is known that spectral correlations exhibit random-matrix universality when the energy difference is less than a characteristic value called the Thouless energy \footnote{The time scale inversely related to the Thouless energy $E_{\rm Th}$ is the Thouless time $T_{\rm Th} \sim E_{\rm Th}^{-1}$.}. More precisely, for $|E_i -E_j| \ll E_{\mt{Th}}$, the \emph{ensemble-averaged} spectral correlator coincides with that of the appropriate Wigner–Dyson ensemble. As a representative example, we consider the Gaussian Unitary Ensemble (GUE), whose two-point spectral correlator is governed by the well-known sine kernel \footnote{For clarity, we begin with explicit results for the GUE and later extend the analysis to more general conclusions without assuming a specific universality class.}: 
 \begin{equation}\label{eq:sine}
    \begin{split}
  & \overline{\langle D(E_i) D(E_j)\rangle}
  \underset{E_i\neq E_j}{\approx} \bar{D}(E_i)\bar{D}(E_j)
  -\frac{\sin^2(e^{S_0}\pi E_{ij}\bar{D}(\bar{E}))}{(e^{S_0}\pi E_{ij})^2} \,.
    \end{split}
\end{equation}
As $E_{ij}\to 0$, the off-diagonal part of this correlator vanishes, explicitly realizing the universal phenomenon of level repulsion.

Combining this universal spectral correlation \eqref{eq:sine} with the particular pole structure \eqref{eq:pole} of the matrix elements, we can explicitly derive the time evolution of the generating function by substituting these two ingredients into the spectral representation \eqref{eq:spectraltwo}. The resulting integral can be evaluated over the complex $E_{ij}$ plane. For example, the integral contour corresponding to the classical contributions is illustrated in Fig.~\ref{fig:pole01}. For a generic value of the regulator $\alpha$ satisfying $\alpha T_{\mathrm{H}} \gg 1$, we find that the time evolution of the generating function is dominated by the following contribution (see Supplemental Material \cite{SM}\nocite{Yang:2018gdb,toappear} for a detailed derivation):
\begin{equation}\label{eq:sloperampplateau}
  \langle\widehat{e^{-\alpha \mC}} \rangle \approx 
\begin{cases}
e^{-\tilde{\alpha} t} + \mathcal{O}(\frac{\talpha }{t \Delta E})\,, \quad \text{slope} \quad \tilde{\alpha} t \sim 1  \,,\\[1em]
\dfrac{2t}{\tilde{\alpha} \, T_{\mt{H}}^2} + \mathcal{O} (e^{-\talpha t}) \,,\quad \text{ramp} \quad  t <   T_{\mt{H}}  \,,\\[1em]
 \dfrac{2}{\tilde{\alpha} \, T_{\mt{H}}} + \mathcal{O}(e^{-\talpha (t-T_{\mt{H}})})\,,\quad \text{plateau} \quad  t >   T_{\mt{H}}  \,. \\
\end{cases}
\end{equation}
A similar behavior has been explicitly obtained for geodesic lengths and time shifts in JT gravity~\cite{geodesic}. Interestingly, as illustrated in \cite{geodesic}, the linear ramp region would gradually disappear as the control parameter $\alpha $ decreases. The non-perturbative expectation value of quantum (holographic) complexity $\mC$ is derived from its regularized generating function by 
\begin{equation}\label{eq:definederivative}
  {\mC}  := - \lim_{\alpha \to 0}  \partial_\alpha \left(\langle\widehat{e^{-\alpha \mC}} \rangle_{\text{reg}}  \right)\,,
\end{equation}
where we perform the regularization according to Eq.~\eqref{eq:complete}. By explicitly using the pole structure given in Eq.~\eqref{eq:pole} and the sine kernel \eqref{eq:sine}, we obtain a simpler time evolution:
\begin{equation}\label{eq:Vgenvalues}
  {\mC}  \approx \text{const} + 
\begin{cases}
      M\,t\left(1-\frac{t}{T_{\mt{H}}}+\frac{t^2}{3 T_{\mt{H}}^2}\right) \,, (t < T_{\mt{H}})  \\[1em]
     \frac{1}{3} MT_{\mt{H}} \,, \quad (t \ge T_{\mt{H}}) \,,
 \end{cases} \,. 
\end{equation}
This result clearly exhibits a long-time linear growth and saturates on the plateau after the Heisenberg time $T_{\mt{H}}$. We further note that the classical linear growth rate $M$ is also determined by the location of the pole, \ie $E_{ij}= \pm i \, M \alpha$. The linear growth rate $M$ can be interpreted as the {\it momentum} of the associated complexity measures, see \eg \cite{Bernamonti:2019zyy,Bernamonti:2020bcf,Belin:2021bga,Belin:2022xmt}. From the viewpoint of holographic complexity, the coefficient $M$ is typically related to the ADM mass of AdS black holes.

\begin{figure}[h]
	\centering	\includegraphics[width=3.37in]{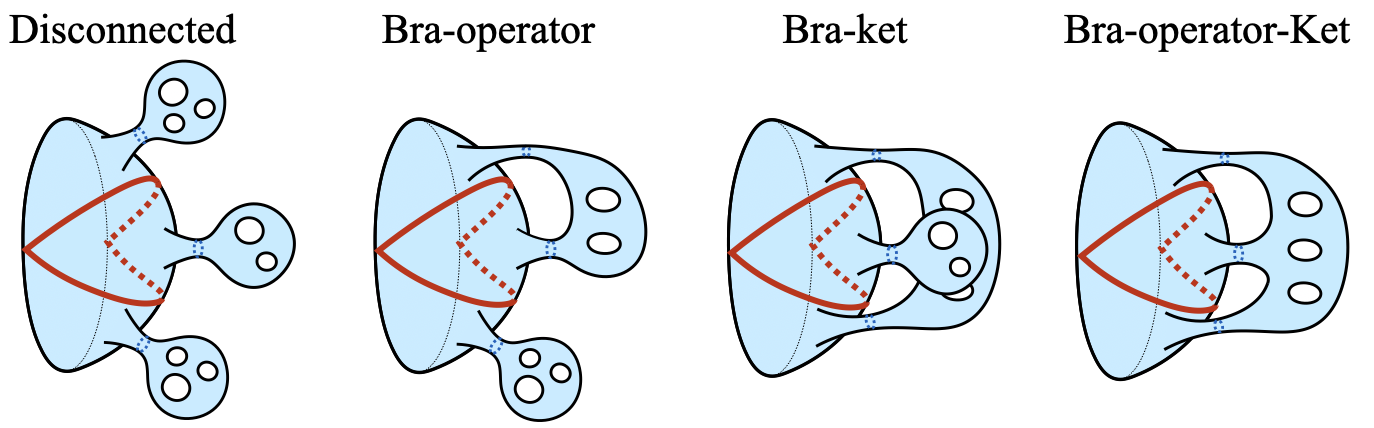}
	\caption{Schematic illustration of Euclidean wormhole saddles suggested by the decomposition of the three-point spectral correlator in Eq.~\eqref{eq:DDD}. Each diagram represents a distinct topology of connectivity among the three sectors (bra, operator insertion, and ket).}\label{fig:wormholes}
\end{figure} 

\vspace{4 pt}
\noindent \emph{4.~\textbf{Codimension-zero observables}.—}The infinitely many codimension-one holographic complexity measures discussed above capture only a limited amount of spectral information, as they exclusively involve two-point spectral correlation functions. In the spirit of the complexity=anything proposal, the most general class of holographic complexity measures is associated with extremal codimension-zero bulk subregions $\mathcal{M}$. Such regions are bounded by future and past spacelike hypersurfaces $\Sigma_{\pm}$, as shown in Fig.~\ref{fig:CAnything}. As the second part of this work, we propose an extension of the previous analysis to codimension-zero holographic complexity $C_{\rm{Any}}$, based on the specific pole structure in Eq.~\eqref{eq:pole}. 

The key point developed in~\cite{Belin:2021bga,Belin:2022xmt} to demonstrate the linear growth of codimension-zero holographic complexity is that the bulk quantity defined within the subregion $\mathcal{M}$ can be reduced to contributions on its future and past boundaries $\Sigma_\pm$ by applying Stokes' theorem. Consequently, within the complexity=anything framework, the codimension-zero holographic complexity can be decomposed as the sum of two codimension-one gravitational observables defined on $\Sigma_{\pm}$, namely 
\begin{equation}
 \mC_{\rm Any}= \mC_+ (\Sigma_+) + \mC_-(\Sigma_-)  \,.
\end{equation}
Motivated by this decomposition, we propose the generating function for codimension-zero complexity as 
\begin{equation}
\langle e^{-\alpha_+ \mC_+ -\alpha_- \mC_-} \rangle:=  \langle \text{TFD} | \widehat{e^{-\alpha_+ \mC_+ }} \, \widehat{e^{-\alpha_- \mC_- }}  | \text{TFD} \rangle \,.
\end{equation}
Its corresponding spectral representation reads
\begin{equation}\label{eq:exppm}
\begin{split}
 &\frac{e^{3S_0}}{Z}\int dE_1dE_2 dE_3 \, e^{-iE_{12}t} \overline{\langle  D(E_1)D(E_2)D(E_3) \rangle}  \\
  &\qquad\quad\times \langle E_1 | \widehat{e^{-\alpha_+ \mC_+ }}| E_3 \rangle  \times  \langle E_3 | \widehat{e^{-\alpha_- \mC_-}}| E_2 \rangle   \,.
\end{split}
\end{equation}
where we insert an identity operator~\eqref{eq:Identity} expanded in energy eigenstates. A crucial distinction between generalized volume and codimension-zero observables is that the latter incorporates \emph{ensemble-averaged} three-point spectral correlation functions, thereby encoding richer information about the spectrum. Due to random matrix universality, the spectral correlation functions can be expressed in terms of the Christoffel-Darboux kernel. Remarkably, we can decompose the ensemble-averaged three-point spectral correlator in terms of 
\begin{widetext}
\begin{equation}\label{eq:DDD}
    \begin{split}
       \overline{ \langle D(E_1)D(E_2)D(E_3)\rangle }&=D(E_1)D(E_2)D(E_3)+ \langle D(E_1)D(E_2)D(E_3)\rangle_c \\
       &+\langle D(E_1)\rangle  \, \langle D(E_2)D(E_3)\rangle_c+\langle D(E_2)\rangle  \, \langle D(E_3)D(E_1)\rangle_c+\langle D(E_3)\rangle  \, \langle D(E_2)D(E_1)\rangle_c \,.\\      
    \end{split}
\end{equation}
\end{widetext}
Interestingly, the spectral decomposition in Eq.~\eqref{eq:DDD} naturally \emph{suggests} corresponding Euclidean wormhole contributions in the gravitational path integral. These contributions are organized by different connectivity patterns among the three sectors, as schematically illustrated in Fig.~\ref{fig:wormholes}. Terms involving connected two-point correlators correspond to configurations in which only two sectors are connected. In contrast, fully connected three-point correlators represent topologies linking all three sectors.

Noting that each matrix element appearing in Eq.~\eqref{eq:exppm} exhibits the same characteristic pole located at $E_{13}=\pm iM_+\alpha_+$ and $E_{32}=\pm iM_-\alpha_-$, as in Eq.~\eqref{eq:pole}, we can extend our analysis directly to codimension-zero generating functions. In particular, holographic complexity $\mC_{\rm{Any}}$ derived in the {\it classical} AdS black hole spacetime corresponds to the contribution from disconnected correlation functions (\ie taking $\frac{1}{\GN} \sim e^{S_0} \to \infty$). Hence, it is straightforward to get the time evolution of codimension-zero holographic complexity: 
\begin{equation}\label{eq:CAnyClasscial}
   {\mC}_{\mt{Any}} \big|_{\rm classical} \approx \text{const} +   (M_+ + M_-) \, t  \,, 
\end{equation}
which matches the linear growth derived previously within the complexity=anything proposal~\cite{Belin:2021bga,Belin:2022xmt}.

To incorporate quantum corrections associated with spacetime wormholes at finite $S_0$, we apply the averaged three-point spectral correlation function~\eqref{eq:DDD}. Similarly, we find that the generating function $\langle e^{-\alpha_+ \mC_+ -\alpha_- \mC_-} \rangle$ involved with the universal pole structure defined in Eq.~\eqref{eq:pole} typically exhibits a slope-ramp-plateau structure analogous to the codimension-one case. After performing the same regularization in the $\alpha \to 0$ limit, we show that the time evolution of codimension-zero complexity measures is similar to Eq.~\eqref{eq:Vgenvalues} for the codimension-one case (see Supplemental Material \cite{SM}).

\begin{figure}[t]
	\centering	\includegraphics[width=3.3in]{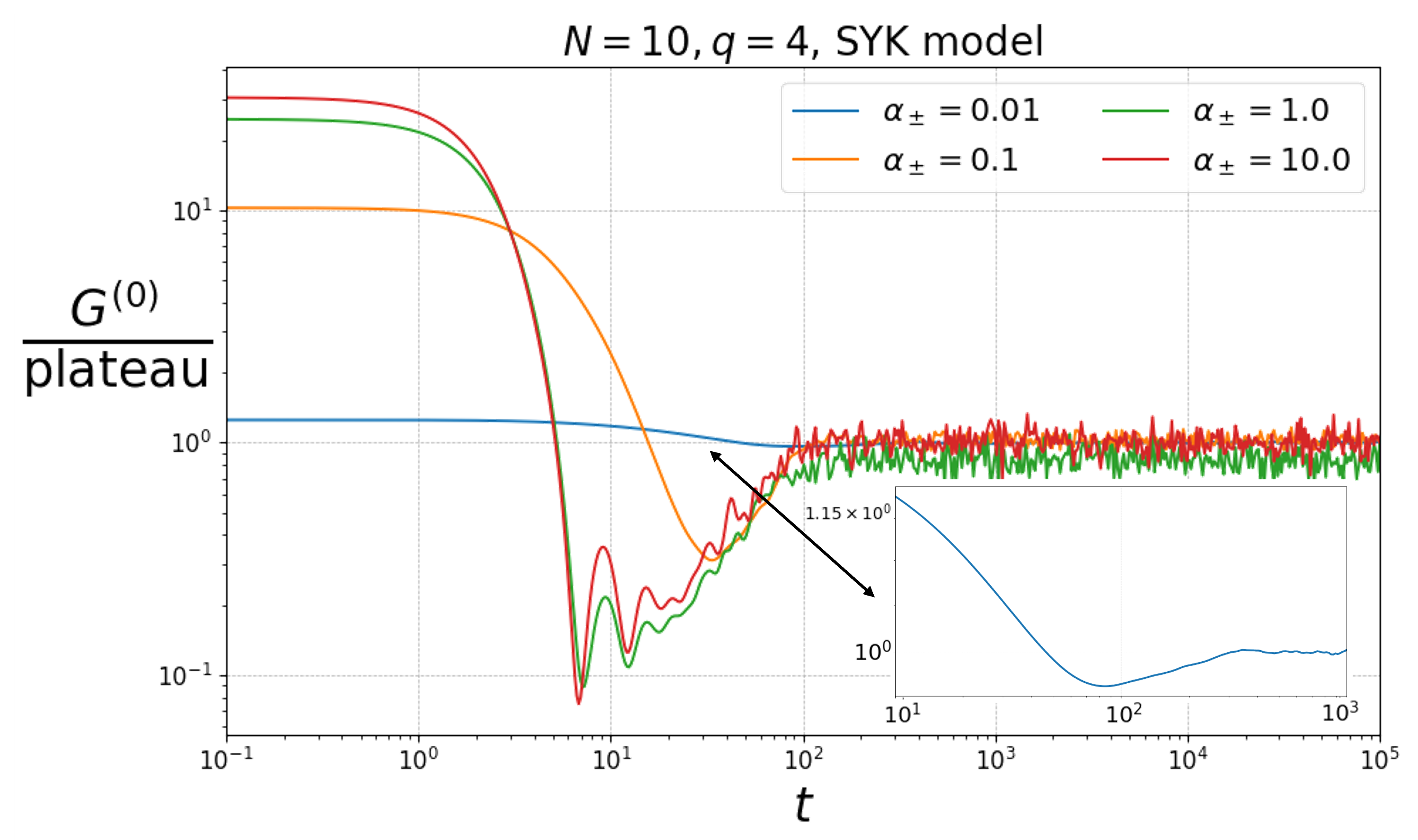}
	\caption{The time evolution of generating functions $\mathrm{G}^{(0)}(\alpha_{\pm},t)$ defined in Eq.~\eqref{eq:Gpm} (rescaled by its plateau value) for the SYK model with $N=10,q=4$. The characteristic linear ramp before the Heisenberg time ($T_{\mt{H}}\sim 10^2$) would be suppressed with decreasing values of $\alpha_{\pm}$.}\label{fig:SYK}
\end{figure}

Although our motivation originates from holographic complexity within the complexity=anything framework, our construction extends naturally to generic quantum systems by defining quantum complexity in the same way. Explicitly, we define universal microscopic generating functions of quantum complexity measures in the microcanonical ensemble \footnote{The canonical version is given by adding a thermal factor $e^{-\frac{\beta}{2}(E_1 +E_2)}$.} as follows: 
\begin{equation}\label{eq:Gpm} 
\mathrm{G}^{(0)}(\alpha_{\pm},t) = \sum_{E_i, E_j, E_k} \frac{\tilde{\alpha}_+ \tilde{\alpha}_-  e^{- i(E_{ik} + E_{kj})t} } {\left( E_{ik}^2 +\tilde{\alpha}_+^2\right)\left( E_{kj}^2 +\tilde{\alpha}_-^2\right)}   \,. 
\end{equation} 
The corresponding codimension-one counterpart has been defined in \cite{geodesic} and is associated with the spectral complexity introduced in \cite{Iliesiu:2021ari}. These generating functions typically exhibit a universal slope-ramp-plateau structure for chaotic quantum systems governed by random matrix universality. As a concrete illustration, numerical results in the SYK model~\cite{Sachdev_1993,KitaevTalks} are shown in Fig.~\ref{fig:SYK}.

\vspace{4 pt}
\noindent \emph{5.~ \textbf{Toward universality in time evolution}.—}
The previous analysis illustrates that the characteristic linear growth and late-time plateau of infinite complexity measures in chaotic systems emerge directly from the specific pole structure described in Eq.~\eqref{eq:pole} and the universal spectral statistics encoded by Eq.~\eqref{eq:sine}. Moreover, we can rigorously identify necessary and sufficient conditions for this universal time evolution using the generating function formalism introduced in eqs.~\eqref{eq:spectraltwo} and \eqref{eq:exppm}. A crucial observation is that, by construction, the time dependence in the spectral representation exclusively arises from the phase factor $e^{-i E_{ij} t}$, originating from the wavefunction of the TFD state. Consequently, upon deforming the integration contour in the complex $E_{ij}$ plane (see Fig.~\ref{fig:pole01}) and applying the residue theorem, the time evolution is thus determined by the residues of poles enclosed by the contour. 
Schematically, we represent this connection as
\begin{equation*}
    \text{Pole Structure} \,\Longleftrightarrow\, \text{Time Evolution}\,.
\end{equation*}
In the classical limit ($e^{S_0}\to\infty$), all quantum corrections vanish, leaving only disconnected spectral correlators. Considering a generic pole located at $E_{ij}=E_{\mt{R}}(\alpha)+ i E_{\mt{I}}(\alpha)$, the residue of the generating function yields a time dependence $e^{-iE_{\mt{R}}t}e^{E_{\mt{I}}t}$. To ensure linear growth of holographic complexity in the classical regime, it is necessary that the matrix element $\langle E_i|\widehat{e^{-\alpha \mC}}|E_j\rangle$ possesses a \emph{first-order pole} satisfying
\begin{equation}
\lim\limits_{\alpha \to 0} E_{\mt R}(\alpha)=0 \,, \,\text{and}\, \,\lim\limits_{\alpha \to 0} E_{\mt I}(\alpha)\sim -M\alpha \to 0 \,. 
\end{equation}
Conversely, we can explicitly verify that no other type of pole (including higher-order poles) can produce the desired linear growth of holographic complexity measures (see Supplemental Material \cite{SM} for a detailed proof).


It is intriguing to note that a corresponding geometric condition appears in the complexity=anything proposal, where the linear growth of holographic complexity follows from the extremization of hypersurfaces $\Sigma$ or codimension-zero bulk subregions $\mathcal{M}$~\cite{Belin:2021bga,Belin:2022xmt}. This analogy motivates us to \emph{conjecture} that the presence of this particular pole structure corresponds holographically to geometric extremization in the bulk spacetime, \ie 
\begin{equation*}
    \text{Extremization} \,\overset{\text{conjecture}}{\Longleftrightarrow}\, \text{Pole at } E_{ij}=-i\tilde{\alpha}\,,
\end{equation*}
thereby establishing a direct link between quantum complexity and its holographic dual description.

Furthermore, we introduce quantum corrections with a finite $S_0$ by incorporating the joint eigenvalue distributions $R_2 (E_i, E_j)$ with a generic Christoffel-Darboux kernel \cite{Eynard:2015aea}. They characterize spectral correlations without contact terms. We find that the necessary and sufficient condition for the late-time plateau of quantum complexity is explicitly given by (see Supplemental Material \cite{SM})
\begin{equation}
 \lim_{E_{ij}=-i\tilde{\alpha}\to 0}\left(\bar{D}(E_i)\bar{D}(E_j)-\text{Kernel}(E_{ij},\bar{E})\right)=0 \,. 
\end{equation}
Remarkably, this condition exactly coincides with spectral level repulsion as shown in Eq.~\eqref{eq:offdiag}. It is worth emphasizing that level repulsion is widely recognized as a hallmark of quantum chaos. Analogous conclusions straightforwardly apply to codimension-zero holographic complexity measures when incorporating three-point spectral correlation functions.

\vspace{4pt}
\noindent \emph{6.~\textbf{Conclusion}.—}
In this Letter, we have presented a universal framework for deriving the time evolution of quantum and holographic complexity based on spectral representations of generating functions. By identifying a characteristic pole structure in the associated matrix elements and incorporating universal spectral correlations from random matrix theory, we have demonstrated that quantum complexity in chaotic quantum systems universally exhibits long-time linear growth followed by a saturation plateau at late times. In particular, we have rigorously established that the presence of this specific pole structure, the finiteness and discreteness of the Hilbert space, and quantum chaos characterized by spectral level repulsion constitute the necessary and sufficient conditions for this universal time evolution of complexity measures. Our results also open several promising directions for future investigation, such as clarifying the relation between holographic complexity measures and Krylov (spread) complexity~\cite{Jian:2020qpp,Balasubramanian:2022tpr,Balasubramanian:2023kwd,Rabinovici:2023yex,Erdmenger:2023wjg,Balasubramanian:2024lqk,Heller:2024ldz}, exploring the switchback effect~\cite{Stanford:2014jda}, extending to quantum complexity measures of mixed states~\cite{Caceres:2019pgf}, and investigating more general complexity operators involving higher-point spectral correlations.

\vspace{8pt}

\begin{acknowledgments}

\noindent \emph{Acknowledgments:}
We are grateful to Vijay Balasubramanian, Johanna Erdmenger, Jonathan Karl, Taishi Kawamoto, Soichiro Mori, Kazumi Okuyama, Andrew Rolph, Masaki Shigemori, Tadashi Takayanagi, Wayne Weng, and Jiuci Xu for valuable discussions and comments, and we extend special thanks to Pawel Caputa, Bowen Chen, Michal Heller, Robert Myers, Jacopo Papalini, Le-Chen Qu, Tim Schuhmann, and Zhuo-Yu Xian for their insightful feedback on the manuscript. MM is supported by JSPS KAKENHI Grant Numbers 24K17044 and 25K00997. SMR is supported by Peking University under startup Grant No. 7101303985. KY and SS are supported by JST SPRING, grant number JPMJSP2125, ``THERS Make New Standards Program for the Next Generation Researchers''. SMR gratefully acknowledges the hospitality of the Instituto de Física Teórica (IFT) and Beijing Normal University during the period when part of this work was carried out. This work was supported by a short-term scientific mission grant from the COST action CA22113 THEORY-CHALLENGES. The authors thank the Yukawa Institute for Theoretical Physics at Kyoto University, where this work was partially done during the YITP-I-25-01 on "Black Hole, Quantum Chaos and Quantum Information".
\end{acknowledgments}

\bibliography{references.bib}

@misc{SM,
  note = {See Supplemental Material at [URL] for details of the spectral representations and time evolution of codimension-one and codimension-zero measures of holographic complexity, and for a general proof of the universal time evolution, which includes Refs.~\cite{Yang:2018gdb,toappear}.}
}

@article{prange1997spectral,
  title = {The Spectral Form Factor Is Not Self-Averaging},
  author = {Prange, R. E.},
  journal = {Phys. Rev. Lett.},
  volume = {78},
  issue = {12},
  pages = {2280--2283},
  numpages = {0},
  year = {1997},
  month = {Mar},
  publisher = {American Physical Society},
  doi = {10.1103/PhysRevLett.78.2280},
  url = {https://link.aps.org/doi/10.1103/PhysRevLett.78.2280}
}

@article{Gharibyan:2018jrp,
    author = "Gharibyan, Hrant and Hanada, Masanori and Shenker, Stephen H. and Tezuka, Masaki",
    title = "{Onset of Random Matrix Behavior in Scrambling Systems}",
    eprint = "1803.08050",
    archivePrefix = "arXiv",
    primaryClass = "hep-th",
    doi = "10.1007/JHEP07(2018)124",
    journal = "JHEP",
    volume = "07",
    pages = "124",
    year = "2018",
    note = "[Erratum: JHEP 02, 197 (2019)]"
}

@article{Sonner:2017hxc,
    author = "Sonner, Julian and Vielma, Manuel",
    title = "{Eigenstate thermalization in the Sachdev-Ye-Kitaev model}",
    eprint = "1707.08013",
    archivePrefix = "arXiv",
    primaryClass = "hep-th",
    doi = "10.1007/JHEP11(2017)149",
    journal = "JHEP",
    volume = "11",
    pages = "149",
    year = "2017"
}

@article{Altland:2020ccq,
    author = "Altland, Alexander and Sonner, Julian",
    title = "{Late time physics of holographic quantum chaos}",
    eprint = "2008.02271",
    archivePrefix = "arXiv",
    primaryClass = "hep-th",
    doi = "10.21468/SciPostPhys.11.2.034",
    journal = "SciPost Phys.",
    volume = "11",
    pages = "034",
    year = "2021"
}

@article{Altland:2022xqx,
    author = "Altland, Alexander and Post, Boris and Sonner, Julian and van der Heijden, Jeremy and Verlinde, Erik P.",
    title = "{Quantum chaos in 2D gravity}",
    eprint = "2204.07583",
    archivePrefix = "arXiv",
    primaryClass = "hep-th",
    doi = "10.21468/SciPostPhys.15.2.064",
    journal = "SciPost Phys.",
    volume = "15",
    number = "2",
    pages = "064",
    year = "2023"
}

@article{Orman:2024mpw,
    author = "Orman, Patrick and Gharibyan, Hrant and Preskill, John",
    title = "{Quantum chaos in the sparse SYK model}",
    eprint = "2403.13884",
    archivePrefix = "arXiv",
    primaryClass = "hep-th",
    doi = "10.1007/JHEP02(2025)173",
    journal = "JHEP",
    volume = "02",
    pages = "173",
    year = "2025"
}

@article{Gautason:2025ryg,
    author = "Gautason, Fri{\dh}rik Freyr and Mohan, Vyshnav and Thorlacius, L{\'a}rus",
    title = "{Late-time saturation of black hole complexity}",
    eprint = "2502.17179",
    archivePrefix = "arXiv",
    primaryClass = "hep-th",
    doi = "10.1007/JHEP08(2025)056",
    journal = "JHEP",
    volume = "08",
    pages = "056",
    year = "2025"
}

@article{Alishahiha:2022nhe,
    author = "Alishahiha, Mohsen",
    title = "{On quantum complexity}",
    eprint = "2209.14689",
    archivePrefix = "arXiv",
    primaryClass = "hep-th",
    doi = "10.1016/j.physletb.2023.137979",
    journal = "Phys. Lett. B",
    volume = "842",
    pages = "137979",
    year = "2023"
}

@article{Alishahiha:2022anw,
    author = "Alishahiha, Mohsen and Banerjee, Souvik",
    title = "{A universal approach to Krylov state and operator complexities}",
    eprint = "2212.10583",
    archivePrefix = "arXiv",
    primaryClass = "hep-th",
    doi = "10.21468/SciPostPhys.15.3.080",
    journal = "SciPost Phys.",
    volume = "15",
    number = "3",
    pages = "080",
    year = "2023"
}

@article{Haferkamp:2021uxo,
    author = "Haferkamp, Jonas and Faist, Philippe and Kothakonda, Naga B. T. and Eisert, Jens and Halpern, Nicole Yunger",
    title = "{Linear growth of quantum circuit complexity}",
    eprint = "2106.05305",
    archivePrefix = "arXiv",
    primaryClass = "quant-ph",
    doi = "10.1038/s41567-022-01539-6",
    journal = "Nature Phys.",
    volume = "18",
    number = "5",
    pages = "528--532",
    year = "2022"
}

@article{Rabinovici:2025otw,
    author = "Rabinovici, Eliezer and S{\'a}nchez-Garrido, Adri{\'a}n and Shir, Ruth and Sonner, Julian",
    title = "{Krylov Complexity}",
    eprint = "2507.06286",
    archivePrefix = "arXiv",
    primaryClass = "hep-th",
    reportNumber = "CERN-TH-2025-128",
    month = "7",
    year = "2025",
    journal = "",
}

@article{Bhattacharjee:2022ave,
    author = "Bhattacharjee, Budhaditya and Nandy, Pratik and Pathak, Tanay",
    title = "{Krylov complexity in large q and double-scaled SYK model}",
    eprint = "2210.02474",
    archivePrefix = "arXiv",
    primaryClass = "hep-th",
    reportNumber = "YITP-22-106",
    doi = "10.1007/JHEP08(2023)099",
    journal = "JHEP",
    volume = "08",
    pages = "099",
    year = "2023"
}

@article{Lin:2022rbf,
    author = "Lin, Henry W.",
    title = "{The bulk Hilbert space of double scaled SYK}",
    eprint = "2208.07032",
    archivePrefix = "arXiv",
    primaryClass = "hep-th",
    doi = "10.1007/JHEP11(2022)060",
    journal = "JHEP",
    volume = "11",
    pages = "060",
    year = "2022"
}

@article{Ambrosini:2024sre,
    author = "Ambrosini, Marco and Rabinovici, Eliezer and S{\'a}nchez-Garrido, Adri{\'a}n and Shir, Ruth and Sonner, Julian",
    title = "{Operator K-complexity in DSSYK: Krylov complexity equals bulk length}",
    eprint = "2412.15318",
    archivePrefix = "arXiv",
    primaryClass = "hep-th",
    reportNumber = "CERN-TH-2025-040",
    doi = "10.1007/JHEP08(2025)059",
    journal = "JHEP",
    volume = "08",
    pages = "059",
    year = "2025"
}

@article{Baiguera:2025dkc,
    author = "Baiguera, Stefano and Balasubramanian, Vijay and Caputa, Pawel and Chapman, Shira and Haferkamp, Jonas and Heller, Michal P. and Halpern, Nicole Yunger",
    title = "{Quantum complexity in gravity, quantum field theory, and quantum information science}",
    eprint = "2503.10753",
    archivePrefix = "arXiv",
    primaryClass = "hep-th",
    reportNumber = "YITP-25-39",
    doi = "10.1016/j.physrep.2025.11.001",
    journal = "Phys. Rept.",
    volume = "1159",
    pages = "1--77",
    year = "2026"
}

@article{Caputa:2023vyr,
    author = "Caputa, Pawel and Magan, Javier M. and Patramanis, Dimitrios and Tonni, Erik",
    title = "{Krylov complexity of modular Hamiltonian evolution}",
    eprint = "2306.14732",
    archivePrefix = "arXiv",
    primaryClass = "hep-th",
    doi = "10.1103/PhysRevD.109.086004",
    journal = "Phys. Rev. D",
    volume = "109",
    number = "8",
    pages = "086004",
    year = "2024"
}

@article{Caputa:2021sib,
    author = "Caputa, Pawel and Magan, Javier M. and Patramanis, Dimitrios",
    title = "{Geometry of Krylov complexity}",
    eprint = "2109.03824",
    archivePrefix = "arXiv",
    primaryClass = "hep-th",
    doi = "10.1103/PhysRevResearch.4.013041",
    journal = "Phys. Rev. Res.",
    volume = "4",
    number = "1",
    pages = "013041",
    year = "2022"
}

@article{Nandy:2024htc,
    author = "Nandy, Pratik and Matsoukas-Roubeas, Apollonas S. and Mart{\'\i}nez-Azcona, Pablo and Dymarsky, Anatoly and del Campo, Adolfo",
    title = "{Quantum dynamics in Krylov space: Methods and applications}",
    eprint = "2405.09628",
    archivePrefix = "arXiv",
    primaryClass = "quant-ph",
    reportNumber = "RIKEN-iTHEMS-Report-24",
    doi = "10.1016/j.physrep.2025.05.001",
    journal = "Phys. Rept.",
    volume = "1125-1128",
    pages = "1--82",
    year = "2025"
}

@article{Balasubramanian:2022dnj,
    author = "Balasubramanian, Vijay and Magan, Javier M. and Wu, Qingyue",
    title = "{Tridiagonalizing random matrices}",
    eprint = "2208.08452",
    archivePrefix = "arXiv",
    primaryClass = "hep-th",
    doi = "10.1103/PhysRevD.107.126001",
    journal = "Phys. Rev. D",
    volume = "107",
    number = "12",
    pages = "126001",
    year = "2023"
}

@article{Hashimoto:2023swv,
    author = "Hashimoto, Koji and Murata, Keiju and Tanahashi, Norihiro and Watanabe, Ryota",
    title = "{Krylov complexity and chaos in quantum mechanics}",
    eprint = "2305.16669",
    archivePrefix = "arXiv",
    primaryClass = "hep-th",
    reportNumber = "KUNS-2967",
    doi = "10.1007/JHEP11(2023)040",
    journal = "JHEP",
    volume = "11",
    pages = "040",
    year = "2023"
}

@article{Erdmenger:2023wjg,
    author = "Erdmenger, Johanna and Jian, Shao-Kai and Xian, Zhuo-Yu",
    title = "{Universal chaotic dynamics from Krylov space}",
    eprint = "2303.12151",
    archivePrefix = "arXiv",
    primaryClass = "hep-th",
    doi = "10.1007/JHEP08(2023)176",
    journal = "JHEP",
    volume = "08",
    pages = "176",
    year = "2023"
}

@article{Rabinovici:2022beu,
    author = "Rabinovici, E. and S\'anchez-Garrido, A. and Shir, R. and Sonner, J.",
    title = "{Krylov complexity from integrability to chaos}",
    eprint = "2207.07701",
    archivePrefix = "arXiv",
    primaryClass = "hep-th",
    doi = "10.1007/JHEP07(2022)151",
    journal = "JHEP",
    volume = "07",
    pages = "151",
    year = "2022"
}

@article{Nandy:2024zcd,
    author = "Nandy, Pratik",
    title = "{Tridiagonal Hamiltonians modeling the density of states of the double-scaled SYK model}",
    eprint = "2410.07847",
    archivePrefix = "arXiv",
    primaryClass = "hep-th",
    reportNumber = "RIKEN-iTHEMS-Report-24",
    doi = "10.1007/JHEP01(2025)072",
    journal = "JHEP",
    volume = "01",
    pages = "072",
    year = "2025"
}

@article{Antonini:2024yif,
    author = "Antonini, Stefano and Balasubramanian, Vijay and Bao, Ning and Cao, ChunJun and Chemissany, Wissam",
    title = "{Non-isometry, state dependence and holography}",
    eprint = "2411.07296",
    archivePrefix = "arXiv",
    primaryClass = "hep-th",
    doi = "10.1007/JHEP02(2025)150",
    journal = "JHEP",
    volume = "02",
    pages = "150",
    year = "2025"
}

@article{Akers:2021fut,
    author = "Akers, Chris and Penington, Geoff",
    title = "{Quantum minimal surfaces from quantum error correction}",
    eprint = "2109.14618",
    archivePrefix = "arXiv",
    primaryClass = "hep-th",
    doi = "10.21468/SciPostPhys.12.5.157",
    journal = "SciPost Phys.",
    volume = "12",
    number = "5",
    pages = "157",
    year = "2022"
}

@article{Miyaji:2025ucp,
    author = "Miyaji, Masamichi and Mori, Soichiro and Okuyama, Kazumi",
    title = "{Finite N bulk Hilbert space in ETH matrix model for double-scaled SYK. Null states, state-dependence and Krylov state complexity}",
    eprint = "2505.13194",
    archivePrefix = "arXiv",
    primaryClass = "hep-th",
    reportNumber = "YITP 25-71",
    doi = "10.1007/JHEP08(2025)084",
    journal = "JHEP",
    volume = "08",
    pages = "084",
    year = "2025"
}

@article{Caceres:2019pgf,
    author = "Caceres, Elena and Chapman, Shira and Couch, Josiah D. and Hernandez, Juan P. and Myers, Robert C. and Ruan, Shan-Ming",
    title = "{Complexity of Mixed States in QFT and Holography}",
    eprint = "1909.10557",
    archivePrefix = "arXiv",
    primaryClass = "hep-th",
    doi = "10.1007/JHEP03(2020)012",
    journal = "JHEP",
    volume = "03",
    pages = "012",
    year = "2020"
}

@article{Sato:2025uli,
    author = "Sato, Masayoshi",
    title = "{What happens due to the baby universe effect in JT gravity?Analysis of correlation functions and ERB length at late time using three approaches}",
    journal = {Progress of Theoretical and Experimental Physics},
    eprint = "2503.04331",
    archivePrefix = "arXiv",
    primaryClass = "hep-th",
    doi = "10.1093/ptep/ptaf073",
    month = "3",
    year = "2025"
}

@article{Hernandez:2020nem,
    author = "Hernandez, Juan and Myers, Robert C. and Ruan, Shan-Ming",
    title = "{Quantum extremal islands made easy. Part III. Complexity on the brane}",
    eprint = "2010.16398",
    archivePrefix = "arXiv",
    primaryClass = "hep-th",
    doi = "10.1007/JHEP02(2021)173",
    journal = "JHEP",
    volume = "02",
    pages = "173",
    year = "2021"
}

@article{Bernamonti:2019zyy,
    author = "Bernamonti, Alice and Galli, Federico and Hernandez, Juan and Myers, Robert C. and Ruan, Shan-Ming and Sim\'on, Joan",
    title = "{First Law of Holographic Complexity}",
    eprint = "1903.04511",
    archivePrefix = "arXiv",
    primaryClass = "hep-th",
    doi = "10.1103/PhysRevLett.123.081601",
    journal = "Phys. Rev. Lett.",
    volume = "123",
    number = "8",
    pages = "081601",
    year = "2019"
}

@article{Bernamonti:2020bcf,
    author = "Bernamonti, Alice and Galli, Federico and Hernandez, Juan and Myers, Robert C. and Ruan, Shan-Ming and Sim\'on, Joan",
    title = "{Aspects of The First Law of Complexity}",
    eprint = "2002.05779",
    archivePrefix = "arXiv",
    primaryClass = "hep-th",
    doi = "10.1088/1751-8121/ab8e66",
    journal = "J. Phys. A",
    volume = "53",
    pages = "29",
    year = "2020"
}

@article{Bueno:2016gnv,
    author = "Bueno, Pablo and Min, Vincent S. and Speranza, Antony J. and Visser, Manus R.",
    title = "{Entanglement equilibrium for higher order gravity}",
    eprint = "1612.04374",
    archivePrefix = "arXiv",
    primaryClass = "hep-th",
    doi = "10.1103/PhysRevD.95.046003",
    journal = "Phys. Rev. D",
    volume = "95",
    number = "4",
    pages = "046003",
    year = "2017"
}

@article{geodesic,
    author = "Miyaji, Masamichi and Ruan, Shan-Ming and Shibuya, Shono and Yano, Kazuyoshi",
    title = "{Non-perturbative overlaps in JT gravity: from spectral form factor to generating functions of complexity}",
    eprint = "2502.12266",
    archivePrefix = "arXiv",
    primaryClass = "hep-th",
    reportNumber = "YITP-25-28",
    doi = "10.1007/JHEP06(2025)251",
    journal = "JHEP",
    volume = "06",
    pages = "251",
    year = "2025"
}

@article{Camargo:2024deu,
    author = "Camargo, Hugo A. and Huh, Kyoung-Bum and Jahnke, Viktor and Jeong, Hyun-Sik and Kim, Keun-Young and Nishida, Mitsuhiro",
    title = "{Spread and spectral complexity in quantum spin chains: from integrability to chaos}",
    eprint = "2405.11254",
    archivePrefix = "arXiv",
    primaryClass = "hep-th",
    reportNumber = "IFT-UAM/CSIC-24-65",
    doi = "10.1007/JHEP08(2024)241",
    journal = "JHEP",
    volume = "08",
    pages = "241",
    year = "2024"
}

@article{Heller:2024ldz,
    author = "Heller, Michal P. and Papalini, Jacopo and Schuhmann, Tim",
    title = "{Krylov spread complexity as holographic complexity beyond JT gravity}",
    eprint = "2412.17785",
    archivePrefix = "arXiv",
    primaryClass = "hep-th",
    month = "12",
  journal = "",
    year = "2024"
}

@article{Louko:2000tp,
    author = "Louko, Jorma and Marolf, Donald and Ross, Simon F.",
    title = "{On geodesic propagators and black hole holography}",
    eprint = "hep-th/0002111",
    archivePrefix = "arXiv",
    doi = "10.1103/PhysRevD.62.044041",
    journal = "Phys. Rev. D",
    volume = "62",
    pages = "044041",
    year = "2000"
}

@article{Balasubramanian:1999zv,
    author = "Balasubramanian, Vijay and Ross, Simon F.",
    title = "{Holographic particle detection}",
    eprint = "hep-th/9906226",
    archivePrefix = "arXiv",
    reportNumber = "HUTP-99-A032, UCSBTH-99-1",
    doi = "10.1103/PhysRevD.61.044007",
    journal = "Phys. Rev. D",
    volume = "61",
    pages = "044007",
    year = "2000"
}

@article{toappear,
    author = "Masamichi Miyaji and Shan-Ming Ruan and Shono Shibuya and Kazuyoshi Yano",
    title = "{in preparation}",
    year = "2026",
    journal = "",
    volume = "",
}

@article{Jian:2020qpp,
    author = "Jian, Shao-Kai and Swingle, Brian and Xian, Zhuo-Yu",
    title = "{Complexity growth of operators in the SYK model and in JT gravity}",
    eprint = "2008.12274",
    archivePrefix = "arXiv",
    primaryClass = "hep-th",
    doi = "10.1007/JHEP03(2021)014",
    journal = "JHEP",
    volume = "03",
    pages = "014",
    year = "2021"
}

@article{Brown:2017jil,
    author = "Brown, Adam R. and Susskind, Leonard",
    title = "{Second law of quantum complexity}",
    eprint = "1701.01107",
    archivePrefix = "arXiv",
    primaryClass = "hep-th",
    doi = "10.1103/PhysRevD.97.086015",
    journal = "Phys. Rev. D",
    volume = "97",
    number = "8",
    pages = "086015",
    year = "2018"
}

@article{Jorstad:2023kmq,
    author = "J\o{}rstad, Eivind and Myers, Robert C. and Ruan, Shan-Ming",
    title = "{Complexity=anything: singularity probes}",
    eprint = "2304.05453",
    archivePrefix = "arXiv",
    primaryClass = "hep-th",
    reportNumber = "YITP-23-41",
    doi = "10.1007/JHEP07(2023)223",
    journal = "JHEP",
    volume = "07",
    pages = "223",
    year = "2023"
}

@article{Belin:2022xmt,
    author = "Belin, Alexandre and Myers, Robert C. and Ruan, Shan-Ming and S\'arosi, G\'abor and Speranza, Antony J.",
    title = "{Complexity equals anything II}",
    eprint = "2210.09647",
    archivePrefix = "arXiv",
    primaryClass = "hep-th",
    reportNumber = "CERN-TH-2022-159; YITP-22-101",
    doi = "10.1007/JHEP01(2023)154",
    journal = "JHEP",
    volume = "01",
    pages = "154",
    year = "2023"
}

@article{Belin:2021bga,
    author = "Belin, Alexandre and Myers, Robert C. and Ruan, Shan-Ming and S\'arosi, G\'abor and Speranza, Antony J.",
    title = "{Does Complexity Equal Anything?}",
    eprint = "2111.02429",
    archivePrefix = "arXiv",
    primaryClass = "hep-th",
    reportNumber = "CERN-TH-2021-181, YITP-22-02",
    doi = "10.1103/PhysRevLett.128.081602",
    journal = "Phys. Rev. Lett.",
    volume = "128",
    number = "8",
    pages = "081602",
    year = "2022"
}

@article{Iliesiu:2021ari,
    author = "Iliesiu, Luca V. and Mezei, M\'ark and S\'arosi, G\'abor",
    title = "{The volume of the black hole interior at late times}",
    eprint = "2107.06286",
    archivePrefix = "arXiv",
    primaryClass = "hep-th",
    reportNumber = "CERN-TH-2021-102",
    doi = "10.1007/JHEP07(2022)073",
    journal = "JHEP",
    volume = "07",
    pages = "073",
    year = "2022"
}

@article{Cotler:2016fpe,
    author = "Cotler, Jordan S. and Gur-Ari, Guy and Hanada, Masanori and Polchinski, Joseph and Saad, Phil and Shenker, Stephen H. and Stanford, Douglas and Streicher, Alexandre and Tezuka, Masaki",
    title = "{Black Holes and Random Matrices}",
    eprint = "1611.04650",
    archivePrefix = "arXiv",
    primaryClass = "hep-th",
    reportNumber = "SU-ITP-16-19, SU-ITP-16/19, YITP-16-124",
    doi = "10.1007/JHEP05(2017)118",
    journal = "JHEP",
    volume = "05",
    pages = "118",
    year = "2017",
    note = "[Erratum: JHEP 09, 002 (2018)]"
}

@article{Saad:2018bqo,
    author = "Saad, Phil and Shenker, Stephen H. and Stanford, Douglas",
    title = "{A semiclassical ramp in SYK and in gravity}",
    eprint = "1806.06840",
    archivePrefix = "arXiv",
    primaryClass = "hep-th",
    month = "6",
    year = "2018",
  journal = "",
}

@article{Saad:2019lba,
    author = "Saad, Phil and Shenker, Stephen H. and Stanford, Douglas",
    title = "{JT gravity as a matrix integral}",
    eprint = "1903.11115",
    archivePrefix = "arXiv",
    primaryClass = "hep-th",
    month = "3",
    year = "2019",
    journal = "",
}

@article{Saad:2019pqd,
    author = "Saad, Phil",
    title = "{Late Time Correlation Functions, Baby Universes, and ETH in JT Gravity}",
    journal = "",
    eprint = "1910.10311",
    archivePrefix = "arXiv",
    primaryClass = "hep-th",
    month = "10",
    year = "2019"
}

@article{Yang:2018gdb,
    author = "Yang, Zhenbin",
    title = "{The Quantum Gravity Dynamics of Near Extremal Black Holes}",
    eprint = "1809.08647",
    archivePrefix = "arXiv",
    primaryClass = "hep-th",
    doi = "10.1007/JHEP05(2019)205",
    journal = "JHEP",
    volume = "05",
    pages = "205",
    year = "2019"
}

@article{Saad:2022kfe,
    author = "Saad, Phil and Stanford, Douglas and Yang, Zhenbin and Yao, Shunyu",
    title = "{A convergent genus expansion for the plateau}",
    eprint = "2210.11565",
    archivePrefix = "arXiv",
    primaryClass = "hep-th",
    doi = "10.1007/JHEP09(2024)033",
    journal = "JHEP",
    volume = "09",
    pages = "033",
    year = "2024"
}

@article{Blommaert:2024ftn,
    author = "Blommaert, Andreas and Chen, Chang-Han and Nomura, Yasunori",
    title = "{Firewalls at exponentially late times}",
    eprint = "2403.07049",
    archivePrefix = "arXiv",
    primaryClass = "hep-th",
    reportNumber = "RIKEN-iTHEMS-Report-24",
    doi = "10.1007/JHEP10(2024)131",
    journal = "JHEP",
    volume = "10",
    pages = "131",
    year = "2024"
}

@article{Brown:2015bva,
    author = "Brown, Adam R. and Roberts, Daniel A. and Susskind, Leonard and Swingle, Brian and Zhao, Ying",
    title = "{Holographic Complexity Equals Bulk Action?}",
    eprint = "1509.07876",
    archivePrefix = "arXiv",
    primaryClass = "hep-th",
    doi = "10.1103/PhysRevLett.116.191301",
    journal = "Phys. Rev. Lett.",
    volume = "116",
    number = "19",
    pages = "191301",
    year = "2016"
}

@article{Brown:2015lvg,
    author = "Brown, Adam R. and Roberts, Daniel A. and Susskind, Leonard and Swingle, Brian and Zhao, Ying",
    title = "{Complexity, action, and black holes}",
    eprint = "1512.04993",
    archivePrefix = "arXiv",
    primaryClass = "hep-th",
    doi = "10.1103/PhysRevD.93.086006",
    journal = "Phys. Rev. D",
    volume = "93",
    number = "8",
    pages = "086006",
    year = "2016"
}

@article{Miyaji:2015woj,
    author = "Miyaji, Masamichi and Numasawa, Tokiro and Shiba, Noburo and Takayanagi, Tadashi and Watanabe, Kento",
    title = "{Distance between Quantum States and Gauge-Gravity Duality}",
    eprint = "1507.07555",
    archivePrefix = "arXiv",
    primaryClass = "hep-th",
    reportNumber = "YITP-15-62, IPMU15-0119",
    doi = "10.1103/PhysRevLett.115.261602",
    journal = "Phys. Rev. Lett.",
    volume = "115",
    number = "26",
    pages = "261602",
    year = "2015"
}

@article{Belin:2018bpg,
    author = "Belin, Alexandre and Lewkowycz, Aitor and S\'arosi, G\'abor",
    title = "{Complexity and the bulk volume, a new York time story}",
    eprint = "1811.03097",
    archivePrefix = "arXiv",
    primaryClass = "hep-th",
    doi = "10.1007/JHEP03(2019)044",
    journal = "JHEP",
    volume = "03",
    pages = "044",
    year = "2019"
}

@article{Miyaji:2016fse,
    author = "Miyaji, Masamichi",
    title = "{Butterflies from Information Metric}",
    eprint = "1607.01467",
    archivePrefix = "arXiv",
    primaryClass = "hep-th",
    doi = "10.1007/JHEP09(2016)002",
    journal = "JHEP",
    volume = "09",
    pages = "002",
    year = "2016"
}

@article{Caputa:2017urj,
    author = "Caputa, Pawel and Kundu, Nilay and Miyaji, Masamichi and Takayanagi, Tadashi and Watanabe, Kento",
    title = "{Anti-de Sitter Space from Optimization of Path Integrals in Conformal Field Theories}",
    eprint = "1703.00456",
    archivePrefix = "arXiv",
    primaryClass = "hep-th",
    reportNumber = "YITP-17-20, IPMU17-0039",
    doi = "10.1103/PhysRevLett.119.071602",
    journal = "Phys. Rev. Lett.",
    volume = "119",
    number = "7",
    pages = "071602",
    year = "2017"
}

@article{Caputa:2017yrh,
    author = "Caputa, Pawel and Kundu, Nilay and Miyaji, Masamichi and Takayanagi, Tadashi and Watanabe, Kento",
    title = "{Liouville Action as Path-Integral Complexity: From Continuous Tensor Networks to AdS/CFT}",
    eprint = "1706.07056",
    archivePrefix = "arXiv",
    primaryClass = "hep-th",
    reportNumber = "YITP-17-65, IPMU17-0091",
    doi = "10.1007/JHEP11(2017)097",
    journal = "JHEP",
    volume = "11",
    pages = "097",
    year = "2017"
}

@article{Balasubramanian:2022tpr,
    author = "Balasubramanian, Vijay and Caputa, Pawel and Magan, Javier M. and Wu, Qingyue",
    title = "{Quantum chaos and the complexity of spread of states}",
    eprint = "2202.06957",
    archivePrefix = "arXiv",
    primaryClass = "hep-th",
    doi = "10.1103/PhysRevD.106.046007",
    journal = "Phys. Rev. D",
    volume = "106",
    number = "4",
    pages = "046007",
    year = "2022"
}

@article{Balasubramanian:2023kwd,
    author = "Balasubramanian, Vijay and Magan, Javier M. and Wu, Qingyue",
    title = "{Quantum chaos, integrability, and late times in the Krylov basis}",
    eprint = "2312.03848",
    archivePrefix = "arXiv",
    primaryClass = "hep-th",
    doi = "10.1103/PhysRevE.111.014218",
    journal = "Phys. Rev. E",
    volume = "111",
    number = "1",
    pages = "014218",
    year = "2025"
}

@article{Akers:2022qdl,
    author = "Akers, Chris and Engelhardt, Netta and Harlow, Daniel and Penington, Geoff and Vardhan, Shreya",
    title = "{The black hole interior from non-isometric codes and complexity}",
    eprint = "2207.06536",
    archivePrefix = "arXiv",
    primaryClass = "hep-th",
    doi = "10.1007/JHEP06(2024)155",
    journal = "JHEP",
    volume = "06",
    pages = "155",
    year = "2024"
}

@article{Blommaert:2022lbh,
    author = "Blommaert, Andreas and Kruthoff, Jorrit and Yao, Shunyu",
    title = "{An integrable road to a perturbative plateau}",
    eprint = "2208.13795",
    archivePrefix = "arXiv",
    primaryClass = "hep-th",
    doi = "10.1007/JHEP04(2023)048",
    journal = "JHEP",
    volume = "04",
    pages = "048",
    year = "2023"
}

@article{Okuyama:2023pio,
    author = "Okuyama, Kazumi and Sakai, Kazuhiro",
    title = "{Spectral form factor in the \ensuremath{\tau}-scaling limit}",
    eprint = "2301.04773",
    archivePrefix = "arXiv",
    primaryClass = "hep-th",
    doi = "10.1007/JHEP04(2023)123",
    journal = "JHEP",
    volume = "04",
    pages = "123",
    year = "2023"
}

@article{Okuyama:2020ncd,
    author = "Okuyama, Kazumi and Sakai, Kazuhiro",
    title = "{Multi-boundary correlators in JT gravity}",
    eprint = "2004.07555",
    archivePrefix = "arXiv",
    primaryClass = "hep-th",
    doi = "10.1007/JHEP08(2020)126",
    journal = "JHEP",
    volume = "08",
    pages = "126",
    year = "2020"
}

@article{Rabinovici:2023yex,
    author = "Rabinovici, E. and S\'anchez-Garrido, A. and Shir, R. and Sonner, J.",
    title = "{A bulk manifestation of Krylov complexity}",
    eprint = "2305.04355",
    archivePrefix = "arXiv",
    primaryClass = "hep-th",
    doi = "10.1007/JHEP08(2023)213",
    journal = "JHEP",
    volume = "08",
    pages = "213",
    year = "2023"
}

@article{Maldacena:2001kr,
    author = "Maldacena, Juan Martin",
    title = "{Eternal black holes in anti-de Sitter}",
    eprint = "hep-th/0106112",
    archivePrefix = "arXiv",
    reportNumber = "NSF-ITP-01-59",
    doi = "10.1088/1126-6708/2003/04/021",
    journal = "JHEP",
    volume = "04",
    pages = "021",
    year = "2003"
}

@article{Susskind:2014rva,
    author = "Susskind, Leonard",
    title = "{Computational Complexity and Black Hole Horizons}",
    eprint = "1403.5695",
    archivePrefix = "arXiv",
    primaryClass = "hep-th",
    doi = "10.1002/prop.201500092",
    journal = "Fortsch. Phys.",
    volume = "64",
    pages = "24--43",
    year = "2016",
    note = "[Addendum: Fortsch.Phys. 64, 44--48 (2016)]"
}

@article{Stanford:2014jda,
    author = "Stanford, Douglas and Susskind, Leonard",
    title = "{Complexity and Shock Wave Geometries}",
    eprint = "1406.2678",
    archivePrefix = "arXiv",
    primaryClass = "hep-th",
    doi = "10.1103/PhysRevD.90.126007",
    journal = "Phys. Rev. D",
    volume = "90",
    number = "12",
    pages = "126007",
    year = "2014"
}

@article{Miyaji:2024ity,
    author = "Miyaji, Masamichi",
    title = "{Non-perturbative discrete spectrum of interior length and timeshift in two-sided black hole}",
    eprint = "2410.20662",
    archivePrefix = "arXiv",
    primaryClass = "hep-th",
    reportNumber = "YITP-24-139",
    doi = "10.1007/JHEP04(2025)190",
    journal = "JHEP",
    volume = "04",
    pages = "190",
    year = "2025"
}

@article{Brown:2019rox,
    author = "Brown, Adam R. and Gharibyan, Hrant and Penington, Geoff and Susskind, Leonard",
    title = "{The Python\textquoteright{}s Lunch: geometric obstructions to decoding Hawking radiation}",
    eprint = "1912.00228",
    archivePrefix = "arXiv",
    primaryClass = "hep-th",
    doi = "10.1007/JHEP08(2020)121",
    journal = "JHEP",
    volume = "08",
    pages = "121",
    year = "2020"
}

@article{Balasubramanian:2024lqk,
    author = "Balasubramanian, Vijay and Magan, Javier M. and Nandi, Poulami and Wu, Qingyue",
    title = "{Spread complexity and the saturation of wormhole size}",
    eprint = "2412.02038",
    archivePrefix = "arXiv",
    primaryClass = "hep-th",
    doi = "10.1103/vpyr-b3fb",
    journal = "Phys. Rev. D",
    volume = "113",
    number = "4",
    pages = "046004",
    year = "2026"
}

@article{Sachdev_1993,
   title={Gapless spin-fluid ground state in a random quantum Heisenberg magnet},
   volume={70},
   ISSN={0031-9007},
   url={http://dx.doi.org/10.1103/PhysRevLett.70.3339},
   DOI={10.1103/physrevlett.70.3339},
   number={21},
   journal={Physical Review Letters},
   publisher={American Physical Society (APS)},
   author={Sachdev, Subir and Ye, Jinwu},
   year={1993},
   month=may, pages={3339–3342}
}

@misc{KitaevTalks,
  author       = {Kitaev, Alexei},
  title        = {A simple model of quantum holography},
  howpublished = {Talks at KITP, April 7, 2015 and May 27, 2015},
  note         = {Available at: \url{http://online.kitp.ucsb.edu/online/entangled15/kitaev/}},
}

@article{Griguolo:2023jyy,
    author = "Griguolo, Luca and Papalini, Jacopo and Russo, Lorenzo and Seminara, Domenico",
    title = "{The resurgence of the plateau in supersymmetric $ \mathcal{N} $ = 1 Jackiw-Teitelboim gravity}",
    eprint = "2310.06768",
    archivePrefix = "arXiv",
    primaryClass = "hep-th",
    doi = "10.1007/JHEP06(2024)168",
    journal = "JHEP",
    volume = "06",
    pages = "168",
    year = "2024"
}

@article{Eynard:2015aea,
    author = "Eynard, Bertrand and Kimura, Taro and Ribault, Sylvain",
    title = "{Random matrices}",
    eprint = "1510.04430",
    archivePrefix = "arXiv",
    primaryClass = "math-ph",
    month = "10",
    year = "2015",
   journal = ""
}

@article{Iliesiu:2024cnh,
    author = "Iliesiu, Luca V. and Levine, Adam and Lin, Henry W. and Maxfield, Henry and Mezei, M\'ark",
    title = "{On the non-perturbative bulk Hilbert space of JT gravity}",
    eprint = "2403.08696",
    archivePrefix = "arXiv",
    primaryClass = "hep-th",
    doi = "10.1007/JHEP10(2024)220",
    journal = "JHEP",
    volume = "10",
    pages = "220",
    year = "2024"
}

\onecolumngrid
\begin{center}
{\large \bf{End Matter}}
\end{center}
\twocolumngrid

\vspace{4pt}
\noindent \emph{\textbf{Complete and Orthonormal Basis}.—}
In terms of the energy eigenbasis $|E_i\rangle$, fixed–$\mC$ states can be formally defined as
\begin{equation}
|\mC \rangle = e^{S_0} \int dE \, D(E) \phi_{E}(\mC) |E\rangle \,. 
\end{equation}
All information about the state $|\mC\rangle$ is encoded in the wavefunction
\begin{equation}
\phi_{E}(\mC) \equiv \langle E | \mC \rangle \,. 
\end{equation}
Using this representation, the matrix element entering the generating function of complexity can be defined as
\begin{equation}\label{eq:pole02}
\begin{split}
&\langle E_i | \widehat{e^{-\alpha \mC }}  | E_j \rangle  \equiv  \int d\mC  \, \phi_{E_i}(\mC)\phi^\ast_{E_j}(\mC)  \, e^{-\alpha \mC } \,.  \\
\end{split}
\end{equation}
As shown in the main text, the characteristic pole structure in Eq.~\eqref{eq:pole} is both necessary and sufficient for linear growth of holographic complexity in the semiclassical regime. This particular pole structure is realized by an infinite class of operators, including the geodesic length operator in JT gravity~\cite{Miyaji:2024ity} and the chord-number operator in the double-scaled SYK model~\cite{Miyaji:2025ucp}. The latter is also equivalent to the Krylov spread complexity operator. From the perspective of the wavefunction $\phi_E(\mC)$, the origin of this pole can be traced to its characteristic oscillatory behavior. In particular, for $E_i\simeq E_j$ one finds 
\begin{equation}\label{eq:wavefunction}
    \phi_{E_i}(\mC)\phi^\ast_{E_j}(\mC) \underset{E_i\approx E_j}{\approx} \frac{1}{e^{S_0} \pi \bar{D}(\bar{E}) M} \cos \left( t_\mC E_{ij} \right) \,,
\end{equation}
with $t_\mC \approx \mC/M$. Substituting this form into Eq.~\eqref{eq:pole02} immediately reproduces the pole structure in Eq.~\eqref{eq:pole}.

An alternative approach to computing the expectation value of complexity, introduced in \cite{geodesic}, is based on the squared overlap between a fixed–$\mC$ state and the thermofield double state,
\begin{equation}
\begin{split}
     &P_t(\mC):= \langle  \text{TFD} (t) | \mC \rangle  \langle\mC|\text{TFD}(t)\rangle   \\
     &= \frac{e^{2S_0}}{Z} \int dE_i dE_j \, e^{-iE_{ij}t} \, \phi_{E_i}(\mC)\phi^\ast_{E_j}(\mC) \, \overline{\langle  D(E_i) D(E_j) \rangle}  \,.
     \end{split}
\end{equation}
Using the characteristic wavefunction \eqref{eq:wavefunction}, one finds that the dominant contribution depends only on $|t-t_\mC|$, since
\begin{equation}
e^{-iE_{ij}t} \, \phi_{E_i}(\mC)\phi^\ast_{E_j}(\mC)  \sim e^{\pm i E_{ij} (t-t_\mC)} \,. 
\end{equation}
Crucially, we highlight that this squared overlap incorporates non-perturbative corrections because it is involved with the ensemble-averaged spectral correlator, \ie 
\begin{equation}
  e^{2S_0} \,\overline{ \langle D(E_i)D(E_j)\rangle }=  R_2(E_i, E_j) +  \delta(E_{ij})  e^{S_0}D(E_i) \,. 
\end{equation}
Equipped with this ``probability distribution", the expectation value of generating function and complexity measures can then be formally obtained as
\begin{equation}\label{eq:genC02}
\langle e^{-\alpha \mC } \rangle  = \int  d \mC \, P_t(\mC) \times  e^{-\alpha \mC} \,  \,,
\end{equation}
and 
\begin{equation}\label{eq:defineC02}
\langle \hat{\mC} \rangle  = \int  d \mC \, P_t(\mC) \times  \mC \,  \,,
\end{equation}
respectively. However, this integral (\ie sum over all basis states) generically results in the divergence of $\langle \hat{\mC} \rangle $ due to contributions from $t_\mC\gg T_{\mt{H}}$. In particular, $P_t(\mC)$ develops a plateau for $t_\mC\gg T_{\mt{H}}$, with $P_t(\mC)\simeq \frac{2}{M T_{\mt{H}}}$. This divergence reflects the over-completeness of the original $\mC$ basis, which in turn originates from spacetime wormhole contributions.

In order to not only remove the divergence but also obtain a physically meaningful operator, one must perform a Gram–Schmidt orthogonalization~\cite{Miyaji:2024ity,Miyaji:2025ucp}, yielding an orthonormal basis $|\mC_O\rangle$. The choice of orthogonalization is directly tied to the physical interpretation of the resulting operator $\hat{\mC}_O$ in the classical regime, in close analogy with non-isometric codes~\cite{Akers:2021fut,Akers:2022qdl,Antonini:2024yif}. The operator $\hat{\mC}_O$ is defined in terms of Eq.~\eqref{eq:defineCO}, and its expectation value reads
\begin{equation}
\langle \hat{\mC}_O \rangle  = \sum_{\text{orthonormal basis}}  P_t(\mC_O) \times  \mC_O \,  \,,
\end{equation}
with $\sum_{\mC_O}  P_t(\mC_O) =1$. In the following, we still denote the total dimension of orthonormal $\mC_O$-basis as the partition function $Z$. In the following, we aim to show that the same universal time evolution can be derived by evaluating $\langle \hat{\mC}_O \rangle$, \ie presenting a long-time linear growth and a transition to the plateau at late times.

In the semiclassical regime $t\ll T_{\mathrm H}$, only the disconnected part of the spectral correlator contributes. Using Eq.~\eqref{eq:wavefunction}, we can easily find that the squared overlap is dominated by 
\begin{equation}
\begin{split}
     P_t(\mC) &\approx 
     \frac{2}{\pi M\Delta E}
     \frac{\sin^2\left((t_{\mC}-t)\frac{\Delta E}{2}\right)}{\big(t_{\mC}-t\big)^2} \,,
     \end{split}
\end{equation}
This expression reveals that the TFD state is sharply localized around $t_\mC \approx t$. 
Corrections due to orthogonalization are suppressed by powers of $e^{-S_0}$, so both the wavefunction structure~\eqref{eq:wavefunction} and the pole structure~\eqref{eq:pole} remain valid to leading order. As a result, the expectation value $\langle \hat{\mC}_O \rangle$ reproduces the expected linear growth at early times, namely
\begin{equation}
\langle \hat{\mC}_{O}  \rangle \big|_{t \ll T_{\mt{H}}} \approx  M \,t \,.
\end{equation}

As the evolution approaches the Heisenberg time $t \sim T{\mt{H}}$, the overlap becomes dominated by states with $\mC_O < Mt$, causing the linear growth to slow down. In the late-time regime $t\gg T_{\mathrm H}$, the probability distribution takes the form
\begin{equation}
   P_t(\mC_O) \approx\frac{1}{Z} +\frac{1}{Z}\int dE_idE_j \phi_{E_i}\phi^\ast_{E_j}R_2(E_i,E_j)e^{iE_{ij}t}     \,. 
\end{equation}
The crucial point is that off-diagonal contributions are suppressed at late times due to level repulsion, $R_2(E_i,E_j)\sim \mathcal{O}(|E_{ij}|^\beta)$, with $\beta=2$ for GUE. As a consequence of this suppression, the probability distribution at late times becomes approximately uniform,  
\begin{equation}
P_t(\mC_O) \approx \frac{1}{Z} + \mathcal{O}\left(\frac{1}{t}\right) \,.
\end{equation}
Here, the constant is fixed by the completeness of orthonormal $\mC_O$-basis. Accordingly, the late-time expectation value saturates to a constant, \ie 
\begin{equation}
\langle \hat{\mC}_{O}  \rangle \big|_{t \gg T_{\mt{H}}} \approx \frac{1}{Z}\sum_{\mC_O-\text{basis}} \mC_O \,.
\end{equation}
Finally, we recover the same universal time evolution, characterized by linear growth and late-time saturation, in agreement with the result obtained from the regularization of generating functions in the $\alpha \to 0$ limit.

\setcounter{equation}{0} 
\renewcommand{\theequation}{S\arabic{equation}} 

\titlepage

\setcounter{page}{1}

\begin{center}
{\LARGE Supplemental Material}
\end{center}

\section{Appendix A. Codimension-one Observables and Spectral Two-point Function}\label{sec:appA}

In this appendix, we aim to illustrate  the origin of the universal time evolution of codimension-one holographic complexity $\mC$, along with its associated generating functions $\langle \widehat{e^{-\alpha \mC }} \rangle$, in chaotic quantum systems. The spectral representation of the generating function can be recast as
\begin{equation}\label{eq:spectraltwopoint}
 \langle \widehat{e^{-\alpha \mC }} \rangle \equiv  \langle \text{TFD}(t)|\, \widehat{e^{-\alpha \mC }} \, |\text{TFD}(t) \rangle  = \frac{e^{2S_0}}{Z} \int dE_idE_j \, e^{-iE_{ij}t} \overline{\langle  D(E_i) D(E_j) \rangle}  \times  \langle E_i | \widehat{e^{-\alpha \mC }}| E_j \rangle  \,.
\end{equation}
We note that the matrix element $\langle E_i | \widehat{e^{-\alpha \mC }}| E_j \rangle $ on the boundary Hilbert space $\mathbf{E}_{\mt{L}}\otimes \mathbf{E}_{\mt{R}}$ corresponds to the squared matrix elements $| \langle\mathbf{E}_i| \mathcal{O}_{\mt{L/R}} | \mathbf{E}_j\rangle|^2$ of a special Hermitian operator $\mathcal{O}_{\mt{L}} = \mathcal{O}_{\mt{R}}$, which is symmetric under the exchange $ \mathbf{E}_i \leftrightarrow \mathbf{E}_j$. This symmetry ensures that the expectation value $\langle \widehat{e^{-\alpha \mC}} \rangle$ is real, as expected. To make the scaling with $e^{S_0}$ explicit, we adopt the rescaled density of states $D(E_i)$, as used in the main text. This is related to the original density of states by
\begin{equation}\label{eq:rho}
 \rho(E_i) = e^{S_0} D(E_i) \,. 
\end{equation}
We would like to emphasize that most of the quantities that we will calculate are defined as ensemble-averaged values. For instance,  the microscopic spectral density for a non-perturbatively discrete spectrum is given by $\rho(E) = \sum_i \delta (E-E_i)$. The density of states $\rho(E)$ \eqref{eq:rho}, is defined as an averaged (or coarse-grained) density.

Assuming random matrix universality, the averaged spectral correlations $\langle D(E_i)D(E_j) \rangle$ exhibit a universal structure in the limit $E_i \approx E_j$. For the Gaussian Unitary Ensemble (GUE), the spectral two-point correlation function with small energy separations is given by
\begin{equation}\label{eq:sinekernel}
\begin{split}
 \overline{\langle  D(E_i) D(E_j)  \rangle }&=    \bar{D}(E_i) \bar{D}(E_j) + \langle D(E_i) D(E_j) \rangle_c \\
 &\approx    \bar{D}(E_i) \bar{D}(E_j)     +
    e^{-S_0}\delta(E_i-E_j) \bar{D}(E_i)-\left(\frac{\sin (\pi e^{S_0} \bar{D}(\bar{E})(E_i-E_j))}{e^{S_0}\pi (E_i-E_j) }  \right)^2\,, 
    \end{split}
\end{equation}
with $\bar{E}= \frac{E_i +E_j}{2}$ as the mean energy and $E_{ij}= E_i -E_j$ as the energy difference. The delta function term accounts for contact contributions, whereas the sine kernel encodes non-perturbative effects. Importantly, the sine kernel is a hallmark of the GUE in random matrix theory, reflecting level repulsion and spectral rigidity. It is worth emphasizing that the expression in Eq.~\eqref{eq:sinekernel} is valid only within the universal regime near $E_i \approx E_j$. In particular, we have neglected higher-order corrections in $E_{ij}$ that arise at finite $e^{S_0}$, as these are non-universal and sensitive to microscopic details. Accordingly, throughout our analysis of quantum corrections to complexity, we restrict attention to this universal regime, with energies near the microcanonical center $E_i \sim E_0 \sim E_j$, to isolate universal features that are independent of the detailed microscopic structure of the system.

Before proceeding further, we highlight two important limiting cases relevant to our discussion. First, the contribution corresponding to the classical spacetime geometry arises from the disconnected term. This means all quantum corrections vanish in the classical limit, which we explicitly define as
\begin{equation}
\textbf{Classical limit}: \qquad \frac{1}{\GN} \sim e^{S_0} \to \infty  \,.  
\end{equation}
In this limit, the spectral correlation function simplifies as follows:
\begin{equation}
\lim_{e^{S_0} \to \infty}  \overline{\langle  D(E_i) D(E_j)  \rangle } =    \bar{D}(E_i) \bar{D}(E_j) \,,
\end{equation}
because all connected terms vanish. Specifically, one finds
\begin{equation}\label{eq:classicallimit}
\boxed{\lim_{e^{S_0} \to \infty}  \left(  \delta(E_i-E_j) \bar{D}(\bar{E})-\frac{\sin ^2(\pi e^{S_0}  \bar{D}(\bar{E})E_{ij})}{ e^{S_0} (\pi \, E_{ij})^2 }   \right) =0 \,,}
\end{equation}
which directly follows from the identity
\begin{equation}
\lim _{N \rightarrow \infty} \frac{(\sin (N x))^2}{N x^2} =  \pi \delta(x) .
\end{equation}
Second, we consider the particular contributions arising from the diagonal spectrum, which can be isolated by taking the limit $E_{ij} \to 0$. In this diagonal limit, we have
\begin{equation}\label{eq:levelGUE}
\textbf{Diagonal limit}: \qquad \lim_{E_{i} \to E_j}  \overline{ \langle  D(E_i) D(E_j)  \rangle } = e^{-S_0}\delta(E_i-E_j) \bar{D}(E_i) \,,  
\end{equation}
as a direct consequence of the level repulsion of the GUE. More explicitly, the level repulsion manifests as
\begin{equation}
\text{Level repulsion:} 
\qquad \lim_{ E_{ij}\to 0} \underbrace{\left(   \bar{D}(E_i) \bar{D}(E_j)    
    -\left(\frac{\sin (\pi e^{S_0} \bar{D}(\bar{E})E_{ij})}{e^{S_0}\pi E_{ij} }  \right)^2  \right)}_{\text{joint eigenvalue distribution: } R_2(E_i, E_j)} =0 \,.  
\end{equation}
We emphasize that the Dirac delta function appearing in the spectral correlation function~\eqref{eq:sinekernel} emerges from taking continuous spectra, which we have adopted for simplicity. Later, we will demonstrate that careful regularization of multiple-delta terms is essential to avoid $\delta^n$-type divergences and obtain physical results for a discrete and complete basis.

\subsection{A.1 Generating functions $\langle \widehat{e^{-\alpha \mC}} \rangle $}\label{sec:appA1}
To explicitly evaluate the generating functions of quantum complexity measures as defined in Eq.~\eqref{eq:spectraltwopoint}, knowledge of the specific matrix elements is required. However, such detailed information is generally unavailable, except for simple examples like the geodesic length in JT gravity. Importantly, the central message of this paper is that the universal time evolution of holographic complexity measures in chaotic systems is governed by a particular pole structure around the origin $\alpha \sim E_{ij} \ll 1$ on the complex energy plane, explicitly given by
\begin{tcolorbox}[top=-3pt, bottom=2pt]
\begin{equation}
\textbf{Universal Pole Structure:}\qquad \langle E_i | \widehat{e^{-\alpha \mC }}  | E_j \rangle  \equiv  \int d\mC  \, \phi_{E_i}(\mC)\phi^\ast_{E_j}(\mC)   e^{-\alpha \mC }  \sim \frac{1}{(\tilde{\alpha} + i E_{ij} )(\tilde{\alpha} - i E_{ij})} \,.
\end{equation}
\end{tcolorbox}
where
\begin{equation}
\tilde{\alpha} = M \alpha + \mathcal{O}(\alpha^2) \,.
\end{equation}
The above pole structure can be derived with considering a smooth spectrum $\mC$ and assuming a generic oscillating wavefunction $\phi_{E}(\mC)$. Crucially, we are interested in studying a complete $\mC$-basis satisfying 
\begin{equation}
 \boxed{  \hat{\mathbbm{1}}  =   \sum_{\text{complete basis}}  | \mC \rangle \langle \mC | =  \lim_{\alpha \to 0}  \widehat{e^{-\alpha \mC }} \,. }
\end{equation}
Correspondingly, taking $\alpha \to 0$ limit of the matrix element with respect to a complete $\mC$-basis should lead to 
\begin{equation}
  \lim_{\alpha \to 0}   \langle E_i | \widehat{e^{-\alpha \mC }}  | E_j \rangle =   \langle E_i  | E_j \rangle \,, \qquad (\text{complete basis}) 
\end{equation}
Using the normalization condition of energy eigenstates $|E_i \rangle$, namely 
\begin{equation}\label{eq:EiEj}
 \langle E_i  | E_j \rangle  \equiv \frac{\delta(E_{ij})}{e^{S_0} \bar{D}(E_i)} \,, 
\end{equation}
and recalling the representation of the Dirac delta function as a limit of the sequence of the following functions: 
\begin{equation}
\pi  \,  \delta (E) =  \lim_{\alpha \to 0} \frac{\alpha}{\alpha^2 + E^2} \,, 
\end{equation}
we can explicitly fix the normalization factor of the pole as
\begin{equation}\label{eq:polenorm}
  \boxed{\text{Normalization factor for the pole} \approx \frac{\tilde{\alpha}}{\pi e^{S_0}  \bar{D}(\bar{E}) }   \approx \frac{2}{T_{\mt{H}}} \tilde{\alpha} + \mathcal{O}(\alpha^2) \,,}
\end{equation}
by matching the result from the $\alpha \to 0$ limit with Eq.~\eqref{eq:EiEj}. We emphasize that the subleading terms of order $\alpha^2$ are not fixed by this normalization. These terms affect only the constant contribution of the classical value of complexity, \ie $\mC_{\rm classical} \approx M t + \text{Con}$, which we neglect throughout this paper.

\begin{figure}[h]
	\centering	\includegraphics[width=2.2in]{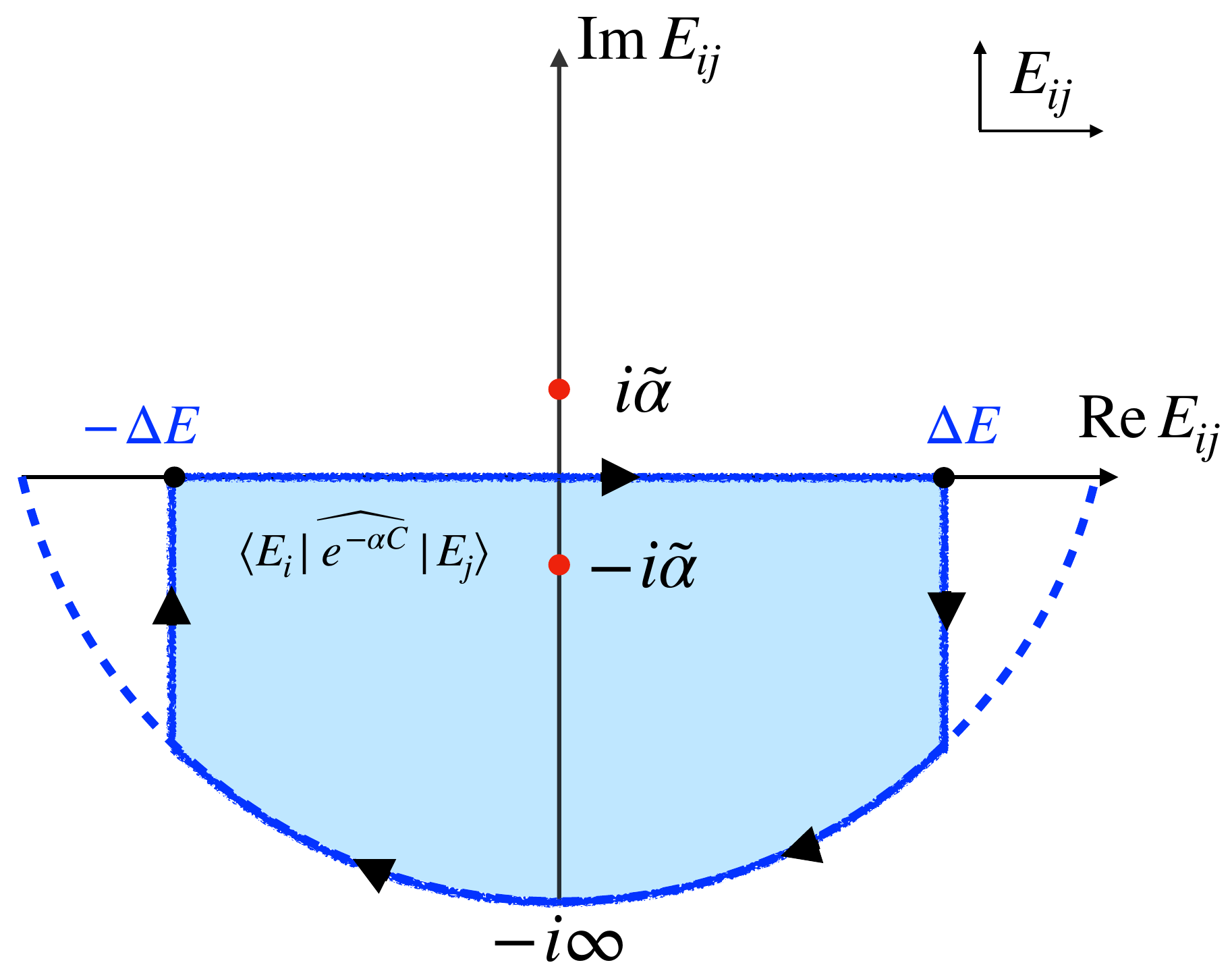}
    \includegraphics[width=4.4in]{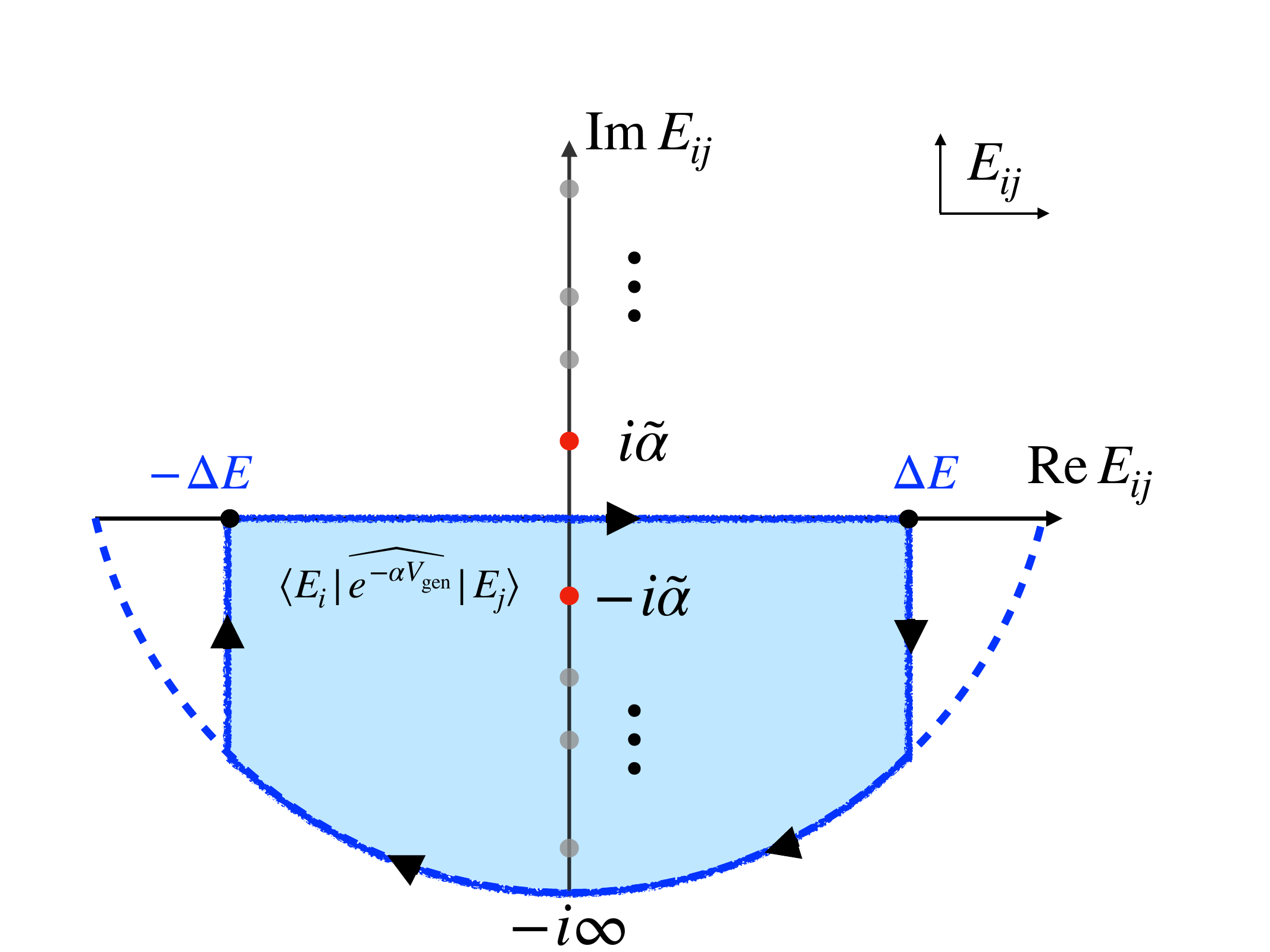}
    \caption{The integral contour defined on the complex plane of $E_{ij}$. Left: the integral contour for the classical part or the late-time regime with $t> T_{\mt{H}}$. Right: two distinct integral contours for different parts in the sine kernel.}\label{fig:contour}
\end{figure}

To explicitly derive the generating functions of complexity $\langle \widehat{e^{-\alpha \mC }}\rangle$ using the spectral representation provided in Eq.~\eqref{eq:spectraltwopoint}, we will separately analyze contributions arising from the disconnected part, the contact term, and the sine kernel with a generic value of $\alpha$, \ie $\alpha T_{\mt{H}} \ll 1$. We will find later that taking $\alpha \to 0$ limit for the complete and discrete basis requires a proper regularization in the treatment of contact terms.

\textbf{01.Disconnected term:}\\
The classical contribution arises from the disconnected correlation function, namely:
\begin{equation}
 \langle \widehat{e^{-\alpha \mC }} \rangle_{\rm classical} \equiv  G_{\rm classical}(\alpha, t) =  \frac{e^{2S_0}}{Z} \int dE_idE_j \, e^{-iE_{ij}t}  D(E_i) D(E_j)\times  \langle E_i | \widehat{e^{-\alpha \mC }}| E_j \rangle  \,.
\end{equation}
We first transform the integral variables from $(E_i, E_j)$ to $(E_{ij}, \bar{E})$, explicitly:
\begin{equation}
\int_{E_0-\Delta E / 2}^{E_0+\Delta E / 2} d E_i d E_j \rightarrow \int_{E_0}^{E_0+\Delta E / 2} d \bar{E} \int_{-\Delta E+2\left(\bar{E}-E_0\right)}^{\Delta E-2\left(\bar{E}-E_0\right)} d E_{ij}+\int_{E_0-\Delta E / 2}^{E_0} d \bar{E} \int_{-\Delta E+2\left(E_0-\bar{E}\right)}^{\Delta E-2\left(E_0-\bar{E}\right)} d E_{ij} \,.
\end{equation}
When performing the integral over the mean energy $\bar{E}$, we only retain the leading contributions and neglect terms suppressed by the small parameter $\frac{\Delta E}{E_0} \ll 1$ characteristic of the microcanonical ensemble. To evaluate the integral conveniently, we analytically continue the integrand to the complex energy plane $E_{ij}$. By appropriately choosing a closed integration contour, we then apply the residue theorem to evaluate the integral on the complex $E_{ij}$ plane, as illustrated in Fig.~\ref{fig:contour}. For $t > 0$, we close the contour in the lower half-plane (as shown in Fig.~\ref{fig:contour}), ensuring exponential decay along the semicircle at infinity. The subleading contributions arising from the finite energy window $\Delta E$ correspond to integrals along the imaginary axis, defined by $E_{ij} = \pm \Delta E + i Y$. Due to symmetry under $E_{ij} \leftrightarrow -E_{ij}$, the imaginary contributions precisely cancel out, as expected. These subleading corrections can be estimated as
\begin{equation}
2 \times \int^{0}_{-\infty} dY \,   \frac{\talpha}{\pi(\Delta E^2 + \talpha^2)} \sin (t \Delta E)e^{t Y}  + \mathcal{O}(\talpha^2) \approx  \frac{2 \talpha \sin (t \Delta E) }{ \pi t \Delta E^2} + \mathcal{O}\left(\frac{1}{(t \Delta E )^2}) \right) + \mathcal{O}(\talpha^2) \,, 
\end{equation}
which are strongly suppressed in the regime of interest, $t \Delta E \gg 1$. Hence, these contributions are negligible when considering either a sufficiently large energy window or sufficiently long times ($t \gg 1$). Consequently, we disregard contributions from the contours along $E_{ij} = \pm i\Delta E + Y$ since we do not address the early-time regime where $t \sim 1/\Delta E$.

Utilizing the explicit Fourier transform:
\begin{equation}
\int^{\infty}_{-\infty}  \, d E_{ij}  \left( \frac{\tilde{\alpha}}{(\tilde{\alpha} + i E_{ij} )(\tilde{\alpha} - i E_{ij})}   \right) e^{-i E_{ij}t} =  \pi e^{- t \tilde{\alpha} }  \,, 
\end{equation}
we thus approximate the classical contribution in the generating function as:
\begin{equation}
\begin{split}
G_{\rm classical}(\alpha, t)  &\approx   \frac{e^{2S_0}}{Z} \int d\bar{E}  \,\bar{\mD}^2  \int dE_{ij} \, \left(  \frac{1}{\pi e^{S_0} \bar{\mD} }   \frac{\tilde{\alpha}}{(\tilde{\alpha} + i E_{ij} )(\tilde{\alpha} - i E_{ij})}   \right) e^{-i E_{ij}t}  \approx  e^{- t \tilde{\alpha} }  \,. 
\end{split}
\end{equation}
Here, we have defined $\bar{\mD} = D(\bar{E}) \approx D(E_0) + \mathcal{O}(\Delta E/ E_0)$. From this explicit result, we conclude that the disconnected contribution corresponds to the ``slope" regime, characterized by exponential decay of the generating function. Importantly, we emphasize that the coefficient of the exponential decay $e^{- t \tilde{\alpha}}$ precisely equals unity, confirming that the linear growth rate of complexity is directly fixed by the relationship $\tilde{\alpha} = M \alpha$.

\textbf{02.Contact term:}\\
The contribution from the contact term yields a constant by definition, as it only involves diagonal matrix elements. Following the same strategy outlined above, it is straightforward to compute
\begin{equation}
G_{\rm delta}(\alpha)= \text{Constant} \approx \frac{2}{T_{\mt{H}}}\int  \, d E_{ij}  \left( \frac{\tilde{\alpha}}{(\tilde{\alpha} + i E_{ij} )(\tilde{\alpha} - i E_{ij})}   \right) e^{-i E_{ij}t} \times  \delta(E_1 - E_2)
= \frac{2}{\tilde{\alpha} T_{\mt{H}}} \,.
\end{equation}
This constant sets the late-time plateau value of the generating function $\langle e^{-\alpha \mC} \rangle$, as illustrated in Fig.~\ref{fig:GeneratingFunctions}.

\textbf{03.Sine kernel}\\
To evaluate the contribution from the sine kernel, we consider a similar Fourier transformation:
\begin{equation}
 \int^\infty_{-\infty} dE_{ij}   \frac{\sin^2 \left( \frac{1}{2} T_{\mt{H}}E_{ij}\right)}{\left( \frac{1}{2} T_{\mt{H}}E_{ij}\right)^2} e^{- i E_{ij} t} \,,
\end{equation}
or, equivalently, apply the residue theorem. The leading contribution from the sine kernel is then found to be:
\begin{equation}\label{eq:Gsine}
\begin{split}
G_{\rm sine}(\alpha, t) 
\approx \begin{cases}
    -\dfrac{2 \talpha (\THH - t)-2 e^{-\talpha t}+e^{\talpha  (t-\THH)}+e^{-\talpha  (t+\THH)}}{\talpha ^2  T_{\mt{H}}^2} \,, \quad \,\, t< T_{\mt{H}}  \\ \, \\
    - \dfrac{4\sinh^2 \left( \frac{1}{2} T_{\mt{H}} \talpha \right) }{\talpha^2 T_{\mt{H}}^2 } e^{- \talpha t} \,,\qquad \qquad  \qquad \quad \qquad t>T_{\mt{H}} \,, \\
    \end{cases} \,. 
\end{split}     
\end{equation}
Notably, this expression reveals the existence of a linear growth regime for $t < T_{\mt{H}}$. While the derivation of this result is straightforward, it is worth emphasizing several subtle and insightful aspects of the sine kernel structure:
\begin{itemize}
   \item Apart from the particular pole at $E_{ij} = \pm i \tilde{\alpha}$, the sine kernel (considered as a unified object) introduces a \textit{removable singularity} ({\it not a pole !}) at the origin $E_{ij} = 0$.
    \item The linear growth regime arises from a \textit{second-order pole} in the constant part of the sine kernel at $E_{ij} = 0$. Decomposing the sine kernel as
\begin{equation}
    2\sin^2(\pi e^{S_0}\bar{\mD} E_{ij}) = 1 - \cos(2\pi e^{S_0}\bar{\mD} E_{ij}) \approx \underbrace{1}_{\to \text{linear ramp}} - \cos \left(T_{\mt{H}} E_{ij} \right) \,,
\end{equation}
we see that the constant term contributes a residue from this second-order pole:
\begin{equation}
    \text{Second-order Pole at } E_{ij}=0 : \qquad \text{Res}\left( \frac{e^{-i E_{ij}t}}{(E_{ij})^2} \right) \bigg|_{E_{ij}=0} \quad \longrightarrow \quad t \,, \qquad \text{for } t < T_{\mt{H}} \,,
\end{equation}
which accounts for the linear ramp of the generating function. This origin of the linear ramp is identical to that seen in the spectral form factor (SFF), as demonstrated by the simple Fourier transform:
\begin{equation}
    -\frac{1}{\pi} \int_{-\infty}^{+\infty} \frac{e^{- i E t }}{E^2} \, dE = |t| \,.
\end{equation}
In JT gravity, the linear ramp of the SFF can be interpreted gravitationally in terms of the double trumpet geometry, corresponding to the connected contribution $Z_{0,2}$ with Euler characteristic $\chi =0$. It would be interesting to interpret the linear ramp of generating functions of complexity from a similar gravitational perspective. Analogous to the SFF, the ramp behavior here may serve as a diagnosis of quantum chaos. In particular, the presence or absence of such a ramp could distinguish between integrable and chaotic systems, as explored in, \eg \cite{geodesic}.

\item The function $G_{\rm sine}(\alpha, t)$ exhibits a phase transition at $t = T_{\mt{H}}$. This can be understood from an alternative decomposition of the sine kernel:
\begin{equation}
    2\sin^2(\pi e^{S_0} \bar{\mD} E_{ij}) \approx 1 - \frac{1}{2} e^{- i T_{\mt{H}} E_{ij}} - \underbrace{\frac{1}{2} e^{+ i T_{\mt{H}} E_{ij}}}_{\text{different contour for } t < T_{\mt{H}}} \,.
\end{equation}
As illustrated in Fig.~\ref{fig:contour}, this decomposition requires two distinct contour prescriptions for $t < T_{\mt{H}}$ and $t > T_{\mt{H}}$. Thus, we may succinctly summarize this phenomenon as:
\begin{equation}
    \boxed{\text{phase transition} = \text{integral contour transition}} \,.
\end{equation}

\item The quantum correction arising from the sine kernel is always exponentially suppressed at late times $t > T_{\mt{H}}$. In this regime, the integral contour is fixed in the lower half-plane (see the left plot in Fig.~\ref{fig:contour}), and the dominant contribution comes from the simple pole at $E_{ij} = -i \tilde{\alpha}$. The associated residue behaves as
\begin{equation}
    \text{Simple Pole at } E_{ij} = - i \tilde{\alpha} : \qquad \text{Res}\left( e^{-i E_{ij} t} \right) \big|_{E_{ij} = -i \tilde{\alpha}} \quad \longrightarrow \quad e^{- \tilde{\alpha} t} \,, \qquad \text{for } t > T_{\mt{H}} \,.
\end{equation}
This observation confirms that the behavior of the generating function at late times is always dominated by the contact term, since it is the only contribution that is independent of the exponentially decaying residue. 
\end{itemize}
Finally, we combine the classical contribution and the quantum corrections as
\begin{equation}
G_{\rm quantum} (\alpha, t) := G_{\rm delta}(\alpha) +G_{\rm sine}(\alpha, t) \,,
\end{equation}
to obtain the full generating function $\langle \widehat{e^{-\alpha \mC }}\rangle$ for complexity measures. Its characteristic time evolution exhibits the well-known slope-ramp-plateau structure at a generic value of $\alpha$ ($\alpha T_{\mt{H}}\gg1$), \ie 
\begin{tcolorbox}[top=-3pt, bottom=2pt]
\begin{equation}\label{eq:Gclassical}
\langle \widehat{e^{-\alpha \mC }} \rangle  =G_{\rm classical}(\alpha, t) + G_{\rm quantum} (\alpha, t)
\approx 
\begin{cases}
e^{-\talpha t} + \mathcal{O}(\frac{\talpha }{t \Delta E})\,, \qquad \text{slope at early times} \quad \alpha t \sim 1  \,,\\[1em]
\dfrac{2t}{\talpha \, T_{\mt{H}}^2} + \mathcal{O} (e^{-\talpha t}) \,,\qquad \text{ramp at the middle stage} \quad  t <   T_{\mt{H}}  \,,\\[1em]
 \dfrac{2}{\talpha \, T_{\mt{H}}} + \mathcal{O}(e^{-\talpha (t-T_{\mt{H}})})\,,\qquad \text{plateau at late times} \quad  t >   T_{\mt{H}}  \,, \\
\end{cases}
\end{equation}
\end{tcolorbox}
which is the result quoted in the main text.
It provides the leading approximation in the parameter regime where $\tilde{\alpha}$ is finite and does not compete with the average level spacing, \ie $1/T_{\mt{H}}$. A more explicit realization of this behavior in the context of geodesic length in JT gravity can be found in \cite{geodesic}.

\subsection{A.2 Regularization and $\alpha 
\to 0$ limit}
After computing the generating functions of complexity, one can extract the time evolution of holographic complexity measures by taking the limit $\alpha \to 0 $. However, this procedure is potentially divergent due to the the continuum approximation of the spectrum, as well as the over-completeness of the infinite $\mathcal{C}$-basis. To regulate these divergences and obtain physically meaningful results with respect to a complete and discrete basis, it is essential to carefully treat the terms involving products of Dirac delta functions, \ie those arising from the sum over diagonal contributions in the spectrum. Accordingly, one may define a regularized operator by explicitly subtracting the divergent component from $\widehat{e^{-\alpha \mathcal{C}}}$, thereby yielding a finite and well-defined result.

A simple illustrative example arises from considering the identity operator, defined as
\begin{equation}
 \hat{\mathbbm{1}} : = e^{S_0} \int dE \, D(E) \, |E\rangle \langle E|  \,. 
\end{equation}  
By definition, the \textit{ensemble-averaged} expectation value of the identity operator remains unchanged, namely:
\begin{equation}\label{eq:averageone}
 \langle \text{TFD} |  \hat{\mathbbm{1}} |\text{TFD} \rangle  =\langle \text{TFD} | \text{TFD} \rangle  =  1  \,. 
\end{equation}
However, let us now attempt to evaluate this average by explicitly expanding the spectral two-point correlation function:
\begin{equation}
\begin{split}
 \langle \text{TFD} | \hat{\mathbbm{1}} |\text{TFD} \rangle  &=  \frac{e^{2S_0}}{Z} \int dE_idE_j \, e^{-iE_{ij}t} \langle  D(E_i) D(E_j) \rangle  \times  \langle E_i | \hat{\mathbbm{1}}  | E_j \rangle  \\
&= \frac{e^{2S_0}}{Z} \int dE_idE_j \, e^{-iE_{ij}t} \langle  D(E_i) D(E_j) \rangle  \times  \frac{\delta(E_i-E_j)}{e^{S_0} D(E_i)  }  \,.
\end{split}
\end{equation}
This explicit decomposition introduces a problem because the exact (ensemble-averaged) decomposition of the spectral two-point function already contains a Dirac delta function, 
\begin{equation}
   \overline{ \langle D(E_i)D(E_j)\rangle }= e^{-2S_0} \, \overline{\langle \rho(E_i)\rho(E_j)\rangle } =  e^{-2S_0}\left( R_2(E_i, E_j) +  \delta(E_i-E_j) \rho(E_i) \right)\,. 
\end{equation}
Thus, when this expression is multiplied by another Dirac delta function $\delta(E_i - E_j)$ from the matrix element $\langle E_i | \hat{\mathbbm{1}} | E_j \rangle$, the integrand includes a product of two delta functions $\delta(E_i - E_j)^2$, which is ill-defined as a distribution. 

However, it is important to emphasize that this type of superficial divergence is merely an artifact of working in the continuum limit for simplicity. In reality, the Hilbert space under consideration has a finite dimension when $S_0$ is finite. Specifically, the total number of microcanonical states is given by
\begin{equation}
Z \equiv e^{S_0} \int_{E_0-\Delta E / 2}^{E_0+\Delta E / 2} D(E) d E \approx e^{S_0}  \bar{D}\left(E_0\right) \Delta E  \,,
\end{equation}
which is manifestly finite for any finite $S_0$. Assuming a finite and discrete spectrum,
\begin{equation}
 E_i \in \{ E_1, E_2, \cdots E_N\} \,, 
\end{equation}
the total number of states is given by
\begin{equation}
N =  \sum_{i}^{N} 1 \quad \longleftrightarrow \quad Z : = \int d E \,   e^{S_0} D(E) \approx e^{S_0}  \bar{D}(E) \Delta E  \,. 
\end{equation}
The continuum limit is formally obtained by replacing the discrete sum with
\begin{equation}
   \sum_{i}^N   \quad \longleftrightarrow \quad \int_{\Delta E} \rho(E) d E = \int_{\Delta E} dE  \,e^{S_0} D(E) \,. 
\end{equation}
Additionally, the normalization of energy eigenstates determines the correspondence between the Kronecker delta and the Dirac delta function:
\begin{equation}\label{eq:doubledeltaij}
 \delta_{ij}  \longleftrightarrow \langle E_i | E_j \rangle = \frac{\delta\left(E_i-E_j\right)}{e^{S_0}  \bar{D}(E_i)}   \,. 
\end{equation}
From the perspective of the discrete spectrum, it is clear that multiple summations over diagonal spectrum remain finite. For instance,
\begin{equation}
\sum_{i}^N \delta_{ij} \sum_{j}^N \delta_{ij} \times f(E_i, E_j)  = \left(\sum_{i}^N \delta_{ij} \times f(E_i, E_j) \right) \times N  \,, 
\end{equation}
is well-defined. Translating this to continuous integrals, terms involving products of delta functions must be regularized accordingly. In particular, the extra delta function in $\delta(E_i - E_j)^2$ should be interpreted as the inverse mean level spacing: 
\begin{equation}
\delta(E_i - E_j) \sim \text { mean level spacing } \sim \frac{\Delta E}{N} \approx e^{S_0} D(E_i) \,,
\end{equation}
More explicitly, the product of two delta functions is regularized in terms of 
\begin{tcolorbox}[title={Regularization of multiple product of Dirac delta functions},colbacktitle=red!10!white, colback=blue!15!white,coltitle=red!70!black,top=-4pt, bottom=2pt]
    \begin{equation}\label{eq:twodelta}
    \begin{split}
    \delta(E_i - E_j) \delta(E_i - E_j)     &\Longrightarrow  \delta(E_i - E_j) \times e^{S_0} D(E_i)     
    \end{split}
    \end{equation} 
\end{tcolorbox}
This regularization prescription can be made precise using a smooth test function $f(E_i, E_j)$:
\begin{equation}
\begin{split}
&\int dE_i dE_j  \, f(E_i, E_j) \times \delta(E_i - E_j)\times \delta(E_i - E_j)  \\
&\Rightarrow \int dE_i dE_j  \, f(E_i, E_j) \delta(E_i - E_j) \times e^{S_0} D(E_i) \\
&= \int dE_i \, f(E_i, E_i) \times  e^{S_0} D(E_i) \,.  
\end{split}
\end{equation}
which is the continuous counterpart of the discrete relation in Eq.~\eqref{eq:doubledeltaij}.

As an exercise and consistency check for the regularization scheme, let us revisit the expectation value of the identity operator from Eq.~\eqref{eq:averageone}. First, using the level repulsion condition in Eq.~\eqref{eq:levelGUE},
\begin{equation}
R_2 (E_i, E_j) \times \delta (E_i- E_j) =0  \,,
\end{equation}
we find that
\begin{equation}
\begin{split}
  \langle D(E_i)D(E_j)\rangle \times \delta(E_i - E_j) &= \left( e^{-2S_0} R_2^{(\rho)}(E_i, E_i) +    e^{-S_0} D(E_i) \delta(E_i -E_j) \right)   \delta(E_i -E_j)  \\
  &= e^{-S_0} D(E_i) \delta(E_i -E_j) \delta(E_i -E_j) \,.
 \end{split}
 \end{equation} 
Applying the regularization rule from Eq.~\eqref{eq:twodelta}, we obtain the physically consistent result, \ie 
\begin{equation}
 \boxed{  \langle D(E_i)D(E_j)\rangle \times \delta(E_i - E_j)  \Longrightarrow  D(E_i) D(E_i) \delta(E_i -E_j) \,. }
\end{equation}
This leads directly to the expected normalization:
\begin{equation}\label{eq:doubledelta}
\begin{split}
\langle \text{TFD} | \hat{\mathbbm{1}} |\text{TFD} \rangle   &=  \frac{e^{2S_0}}{Z} \int dE_i dE_j  \, \left( D(E_i) D(E_i) \delta(E_i -E_j) \right)\times \frac{1}{e^{S_0} D(E_i)} \\
&= \frac{1}{Z} \int dE_i \,  e^{S_0}D(E_i) \\
&=1 \,. 
\end{split}
\end{equation}
Next we consider $\alpha\rightarrow 0$ limit of the generating function for a complete basis. In this limit, the matrix element becomes 
\begin{equation}\label{eq:completematrixelement}
\lim_{\alpha \to 0}  \langle E_i | \widehat{e^{-\alpha \mC }}| E_j \rangle   \rightarrow \langle E_i | E_j \rangle  = \frac{\delta(E_i -E_j)}{e^{S_0}\bar{D}(\bar{E})} \,. 
\end{equation}
Since the average level spacing is given by $T_{\mt{H}}^{-1}$, the discreteness of the spectrum becomes significant in this limit. It is therefore necessary to properly regularize the contribution arising from summing over all diagonal terms, \ie $G_{\rm delta}$ term: 
\begin{equation}
\begin{split}
\lim_{\alpha \to 0}  \underbrace{G_{\rm delta}(\alpha)}_{\text{It is a constant}} &\approx \int  \, d E_{ij} \, e^{-i E_{ij}t}  \delta(E_i - E_j) \times  \underbrace{\left[ \frac{\delta(E_i -E_j)}{e^{S_0}D(\bar{E})}  \right]}_{\text{need to be normalized} \to 1 }  \longrightarrow 1 \,. \\
\end{split}
\end{equation}
The regularization of multiple product of Dirac delta functions introduced in Eq.~\eqref{eq:twodelta} indicates that the $\alpha \to 0$ limit of the contact term should be regularized by 
\begin{tcolorbox}[colbacktitle=red!10!white, colback=blue!15!white,coltitle=red!70!black,top=-4pt, bottom=2pt]
\begin{equation}\label{eq:regdelta}
 \lim_{\alpha \to 0}  G_{\rm delta}^{\rm (reg)}(\alpha)  = 1 \,, \qquad (\text{complete basis})  \,.
\end{equation}
\end{tcolorbox}
As a result, the physically well-defined $\alpha \to 0$ limit of the generating function of complexity is given by 
\begin{equation}\label{eq:alphapluszero}
 \boxed{\lim_{\alpha \to 0}\langle \text{TFD} | \widehat{e^{-\alpha \mC }} | \text{TFD} \rangle  \equiv   \lim_{\alpha \to 0} \langle \widehat{e^{-\alpha \mC }} \rangle = \lim_{\alpha \to 0} G_{\rm delta}^{\rm (reg)}(\alpha)  = 1 \,.}  
\end{equation}

Interestingly, all divergences arising in the limit 
$\alpha \to 0$ can be attributed to a single divergent term proportional to $1/\alpha$ because $\alpha$ parameter in the generating function $\langle \widehat{e^{-\alpha \mC }}\rangle$ plays the role of a regulator for the over-complete basis. This observation motivates the definition of a regularized operator designed to subtract unphysical divergences. It is straightforward to notice that the regularized operator should take the form
\begin{equation}
  \boxed{  \widehat{\left(e^{-\alpha \mathcal{C}}\right)}_{\mathrm{reg}} \approx \widehat{e^{-\alpha \mathcal{C}}}-\hat{\mathbb{I}}\left(\frac{2}{\tilde{\alpha} T_{\mt{H}}}-1+\frac{\talpha}{3} T_{\mt{H}}+\mathcal{O}\left(\alpha^2\right)\right) }\,.
\end{equation}
This prescription precisely reproduces the result obtained from the regularized diagonal contribution $ G_{\rm delta}^{\rm (reg)}(\alpha) $, namely 
\begin{equation}
 \lim_{\alpha \to 0} \langle (\widehat{e^{-\alpha \mC }})_{\text{reg}} \rangle =1= \lim_{\alpha \to 0} \langle \widehat{e^{-\alpha \mC }} \rangle_{\text{reg}} := \lim_{\alpha \to 0} \left(  G_{\rm classical}(\alpha, t) + G^{\text{(reg)}}_{\rm delta}(\alpha) +G_{\rm sine}(\alpha, t)\right) \,. 
\end{equation}
The determination of subleading terms in this expansion will be addressed in the following subsection.

\subsection{A.3 Time evolution of holographic complexity measures}
From the generating function of complexity, one can compute the expectation value of the complexity operator via
\begin{equation}\label{eq:defineC}
\boxed{ \langle \hat{C} \rangle \equiv \langle  \text{TFD}(t)|\, \hat{C} \, |\text{TFD}(t) \rangle = \lim_{\alpha \to 0} \left(  - \partial_\alpha   \langle \widehat{e^{-\alpha \mathcal{C}}} \rangle  \right)_{\mathrm{reg}}  \,. } 
\end{equation}
See, for example, \cite{Yang:2018gdb,Iliesiu:2021ari,Heller:2024ldz,geodesic} for previous studies of geodesic length in JT gravity. However, we emphasize that this expression,  as well as the operator $\hat{C}$, is in general ill-defined if one naively uses an over-complete $\mC$-basis. In this work, we assume the existence of a complete basis ${|\mC\rangle}$. 

By carefully taking the $\alpha \to 0$ limit while avoiding divergences due to the continuum approximation, we may extract the time dependence of complexity either by taking the derivative as in Eq.~\eqref{eq:defineC}, or via a series expansion:
\begin{equation}
\boxed{\langle (\widehat{e^{-\alpha \mC }})_{\text{reg}} \rangle  \approx  1 - \alpha  \,  \langle \hat{C} \rangle   + \mathcal{O}(\alpha^2) \,.} 
\end{equation}
Using the regularized contact term, we define its series expansion with $\alpha \sim 0$ as
\begin{equation}
G^{\rm (reg)}_{\rm delta}(\alpha) \approx  1 + \alpha \, \text{Constant} +  \mathcal{O}(\alpha^2)  \,. 
\end{equation}
In the classical limit of Eq.~\eqref{eq:classicallimit}, one finds
\begin{equation}\label{eq:deltareg}
 \lim_{e^{S_0} \to \infty}  G_{\rm quantum} (\alpha , t) = \lim_{e^{S_0} \to \infty}  \left( G_{\rm delta}(\alpha) +G_{\rm sine}(\alpha, t) \right) =0 \,,    
\end{equation}
which leads to a canonical choice for the subleading coefficient:
\begin{equation}
 \lim_{\talpha \to 0} \left( \partial_{\talpha} G^{\rm (reg)}_{\rm delta}(\alpha)  \right) =  -\frac{T_{\mt{H}}}{3} + 0 + \mathcal{O}\left( \left(\frac{1}{S_0}\right)^{n>0} \right)     \,, 
\end{equation}
using the explicit expression of $G_{\rm sine}(\alpha, t)$ that was derived earlier in Eq.~\eqref{eq:Gsine}.

Combining these results, we now compute the time evolution of the holographic or quantum complexity. Focusing on the regime $t < T_{\mt{H}}$, the generating function has been derived as
\begin{equation} 
\begin{split}
\langle \widehat{e^{-\alpha \mC }} \rangle  &=G_{\rm classical}(\alpha, t) + G_{\rm delta}(\alpha) +G_{\rm sine}(\alpha, t) \\
&\approx e^{- t \tilde{\alpha} }  +  \frac{2}{\talpha T_{\mt{H}}} + \left(  -\dfrac{2 \talpha (\THH - t)-2 e^{-\talpha t}+e^{\talpha  (t-\THH)}+e^{-\talpha  (t+\THH)}}{\talpha ^2  T_{\mt{H}}^2} \right)  \,. \\
\end{split}
\end{equation}
After the regularization, we instead have
\begin{equation}
\begin{split}
\langle \widehat{e^{-\alpha \mC }}\rangle_{\text{reg}} &\overset{\alpha \to 0}{\approx} 
  e^{- t \tilde{\alpha} } + \left( 1 - \talpha \frac{T_{\mt{H}}}{3}  + \mathcal{O}(\talpha^2) \right)- \left( e^{- t \tilde{\alpha} } + \talpha \frac{(t- T_{\mt{H}})^3}{3 T_{\mt{H}}^2}  + \mathcal{O}(\talpha^2)\right) \,,\\
&\approx 1  - \talpha \left( t - \frac{t^2}{T_{\mt{H}}} + \frac{t^3}{3 T_{\mt{H}}^2}\right)  +  \mathcal{O}(\talpha^2)  \,. 
\end{split}
\end{equation}
Here, the subleading term defines the complexity growth, with $\tilde{\alpha}\approx M \alpha$ as before. More strictly speaking, the regularization is performed for getting the proper $\alpha \to 0$ limit. 
Performing the derivative as in Eq.~\eqref{eq:defineC}, the time evolution of the complexity is thus derived as 
\begin{tcolorbox}[top=-3pt, bottom=2pt]
\begin{equation}\label{eq:Cone}
\begin{split}
\langle \hat{\mC} \rangle  &= \lim_{\alpha \to 0} \left( -\partial_\alpha  \left( G_{\rm classical}(\alpha, t) + G^{(\rm reg)}_{\rm quantum} (\alpha, t) \right)  \right) \\
&\approx \text{Con} + 
\begin{cases}
  M  \, 
t\left(1-\frac{t}{T_{\mt{H}}}+\frac{t^2}{3 T_{\mt{H}}^2}\right)\,,\,\, (t < T_{\mt{H}})  \\[1em]
   \frac{1}{3} M \,T_{\mt{H}} \,, \quad\qquad\
   \,\, (t \ge T_{\mt{H}}) 
 \end{cases}\,. 
\end{split}
\end{equation}
\end{tcolorbox}
This result gives the expression presented in the main text. Note that the overall additive constant remains undetermined. It originates from both the regularization of $G^{\rm (reg)}_{\rm delta}(\alpha)$ and the subleading term in the normalization factor of the pole discussed in Eq.~\eqref{eq:polenorm}. Nevertheless, the main feature is clear: the expectation value of quantum complexity measures exhibits a long-time linear growth that saturates to a plateau after the Heisenberg time $T{\mt{H}}$. A similar behavior was obtained for JT gravity in \cite{geodesic} by integration from the finite time derivative without explicitly implementing the regularization.
 
Finally, it is important to emphasize that a plateau in the generating function does not automatically imply a similar plateau in the quantum complexity itself. For instance, an expression of the form $e^{-\alpha t} + \text{const}$ still results in a linear growth at late times. The plateau in quantum complexity arises only from a precise cancellation between the linear growth from the classical term and the linear decay from the quantum corrections. This cancellation is governed by the following condition
\begin{equation}
\text{Residue at $E_{ij}= - i \tilde{\alpha}$}  \quad \longrightarrow \quad   \lim_{\tilde{\alpha}  \to 0}  \left( \bar{\mD}^2   - \dfrac{\sinh^ 2(\pi e^{S_0} \bar{\mD} \tilde{\alpha} )}{(e^{S_0}\pi\tilde{\alpha} )^2}  \right) \times  t \, e^{-t \tilde{\alpha}}=0 \,.
\end{equation}
This is equivalent to the property of the original sine kernel with zero energy separation, \ie 
\begin{equation}
  \lim_{E_{ij} \to 0}  \left(  \bar{D}(E_i) \bar{D}(E_j)    
     -\frac{\sin^2(\pi e^{S_0} \bar{D}(\bar{E})E_{ij})}{(e^{S_0}\pi E_{ij})^2}  \right) =0  \,, 
\end{equation}    
which is precisely the manifestation of level repulsion. In a separate appendix, we generalize this conclusion to more generic quantum systems that exhibit quantum chaos.

\section{Appendix B. Codimension-zero Observables and Spectral Three-point Correlation Function}\label{sec:appB}
In this section, we demonstrate the universal time evolution of codimension-zero holographic complexity $\mC_{\rm Any}$ (see Figure \ref{fig:CAnything}) in chaotic systems. The spectral representation of the corresponding generating function is given by
\begin{equation}\label{eq:gen0}
\begin{split}
\langle e^{-\alpha_+ \mC_+ -\alpha_- \mC_-} \rangle &= \frac{e^{3S_0}}{Z}\int dE_1dE_2 dE_3 \, e^{-iE_{12}t} \overline{\langle  D(E_1)D(E_2)D(E_3) \rangle } \times \langle E_1 | \widehat{e^{-\alpha_+ \mC_+ }}| E_3 \rangle \, \langle E_3 | \widehat{e^{-\alpha_- \mC_-}}| E_2 \rangle   \,.
\end{split}
\end{equation}
As in the codimension-one case, the universal time dependence arises from the pole structure of the matrix elements. Specifically,
\begin{equation}
 \langle E_1 | \widehat{e^{-\alpha_+ \mC_+ }}| E_3 \rangle \, \langle E_3 | \widehat{e^{-\alpha_- \mC_-}}| E_2 \rangle   \sim \left( \frac{1}{\pi e^{S_0} D(E_1)} \frac{\tilde{\alpha}_+}{(\tilde{\alpha}_+^2 + (E_{13})^2 )} \right) \times 
 \left( \frac{1}{\pi e^{S_0} D(E_2)} \frac{\tilde{\alpha}_-}{(\tilde{\alpha}_-^2 +(E_{32})^2 )} \right)  \,,
\end{equation}
where the normalization factor has been included as derived in Eq.~\eqref{eq:polenorm}, and the parameters $\tilde{\alpha}_\pm$ satisfy $\tilde{\alpha}_\pm \approx M_\pm \alpha_\pm + \mathcal{O}(\alpha_\pm^2)$.

\subsection{B.1 Spectral three-point correlation function}
In contrast to the codimension-one case, the generating function $\langle e^{-\alpha_+ \mC_+ -\alpha_- \mC_-} \rangle$ involves the spectral three-point correlators. We begin by reviewing relevant aspects of random matrix theory and deriving the universal form of the three-point function $\langle D(E_1)D(E_2)D(E_3)\rangle$, or equivalently $\langle \rho(E_1)\rho(E_2)\rho(E_3)\rangle$. For a comprehensive review on this topic, see \cite{Eynard:2015aea}.

Let $\mathcal{Z}_N$ denote the partition function of the $N \times N$ random matrix model:
\begin{equation}
\mathcal{Z}_N=\int \left(\prod_{i=1}^{N}d{E_i}\, e^{-V\left(E_i\right)}\right) \Delta\left(E_1, \cdots, E_{N}\right)^\beta
\end{equation}
where $\Delta(E_1, \cdots, E_N)$ is the Vandermonde determinant. The joint probability distribution of eigenvalues of random matrices is given by
\begin{equation}\label{eq:matrixR}
    R_n\left(E_1, \cdots E_{n}\right)=\frac{1}{\mathcal{Z}_N} \frac{N!}{(N-n)!}\prod_{i=1}^{n}e^{-V(E_i)}\int \left(\prod_{i=n+1}^{N}d{E_i}\, e^{-V\left(E_i\right)}\right) \Delta\left(E_1, \cdots E_{N}\right)^\beta\,.
\end{equation}
In terms of the Christoffel–Darboux (CD) kernel $K(E_i, E_j)$, the $n$-point joint eigenvalue distribution functions simplify to
\begin{equation}
R_n\left(E_1, \cdots E_{k}\right)=\det_{i, j=1, \cdots n} K\left(E_i, E_j\right) \,,
\end{equation}
with the diagonal part of the kernel defining the eigenvalue density, namely 
\begin{equation}
     R_1 (K_i)  = K\left(E_i,E_i\right)  = \rho (E_i)  \,.
\end{equation}
The first two joint eigenvalue distributions read
\begin{equation}
    \begin{split}
    R_2 (K_i, K_j)  &=  K\left(E_i, E_i\right) K\left(E_j, E_j\right) - K\left(E_i, E_j\right)K\left(E_j, E_i\right) = \rho_i \rho_j - (K_{ij})^2 \,, 
    \end{split}
\end{equation}
and 
\begin{equation}
\begin{split}
R_3\left(E_i, E_j, E_k\right) &=\operatorname{det}\left(\begin{array}{lll}
K\left(E_i, E_i\right) & K\left(E_i, E_j\right) & K\left(E_i, E_k\right) \\
K\left(E_j, E_i\right) & K\left(E_j, E_j\right) & K\left(E_j, E_k\right) \\
K\left(E_k, E_i\right) & K\left(E_k, E_j\right) & K\left(E_k, E_k\right)
\end{array}\right) \\
&= \rho_i \rho_j \rho_k + 2 K_{ij}K_{jk}K_{ik} - \rho_i (K_{jk})^2 -\rho_j (K_{ik})^2  -\rho_k (K_{ij})^2 \,,   
\end{split}
\end{equation}
To simplify some expressions, we have used the abbreviations:
\begin{equation}
K_{ij}= K(E_i, E_j) \,, \qquad \rho_i =K_{ii}= \rho(E_i)\,, \qquad \delta_{i,j}= \delta(E_i -E_j) \,. 
\end{equation} 
Joint eigenvalue distributions or spectral correlation functions play a central role in diagnosing quantum chaos, as they encode statistical information about the eigenvalue spectrum. In particular, one hallmark of chaotic quantum systems is the phenomenon of level repulsion: 
\begin{equation}
    \boxed{\textbf{Level repulsion: }  
    \text{the probability of two nearby levels being arbitrarily close is suppressed.}}
\end{equation}
This manifests in the vanishing of $n$-point spectral correlators when two arguments coincide:
\begin{equation}
  \boxed{  R_n (E_1, \cdots E_i,\cdots E_i, \cdots E_n) =0  \,. }
\end{equation}
To characterize spectral correlations more precisely, one typically studies the $n$-point correlation functions of the density of states $\rho(E)$. These are typically expressed in terms of the smooth correlation functions $R_n(E_1, \dots, E_n)$, which encode non-trivial statistical correlations among the eigenvalues. However, to account for coincident eigenvalues where two or more energy levels are equal, one must also include contact terms (self-correlation), which involve delta functions reflecting the singular nature of the spectrum at such points. For instance, the spectral two-point function takes the form
\begin{equation}
\begin{split}
   \overline{\langle \rho(E_i)\rho(E_j)\rangle}  &=   R_2 (K_i, K_j)  + \delta(E_i -E_j) \rho(E_i) \,.
\end{split}
\end{equation}
Alternatively, motivated by gravitational interpretations, we can write
\begin{equation}\label{eq:rho1rho2}
     \overline{\langle \rho(E_i)\rho(E_j)\rangle} = \rho(E_i)\rho(E_j)  -    \langle \rho(E_i)\rho(E_j)\rangle_c \,,  \quad  \text{with}\quad  \langle \rho(E_i)\rho(E_j)\rangle_c = \rho(E_i) \delta(E_i -E_j) - \left( K(E_i,E_j) \right)^2\,, 
\end{equation}
where the connected part corresponds to contributions from spacetime wormholes, \ie quantum corrections. In the same spirit, three-point spectral correlation function $\overline{\langle\rho(E_i)\rho(E_j)\rho(E_j)\rangle}$ is 
\begin{equation}\label{eq:rho3}
\begin{split}
    \overline{ \langle \rho(E_i)\rho(E_j)\rho(E_k)\rangle }&=R_3(E_i,E_j,E_k)+\delta(E_j-E_i)\delta(E_k-E_i)R_1(E_i)\\
    &\quad+\delta(E_j-E_k)R_2(E_i,E_j)+\delta(E_k-E_i)R_2(E_j,E_k)+\delta(E_i-E_j)R_2(E_k,E_i)  \,. \\
\end{split}
\end{equation}

In the main text, we rewrite the three-point function in analogy with Eq.~\eqref{eq:rho1rho2}, grouping the connected components such as $\langle \rho(E_i) \rho(E_j) \rangle_c$. Since we use the rescaled density of states $\rho(E)= e^{S_0} D(E)$ to make the dependence on the scale $e^{S_0}$ explicit, 
we rewrite the corresponding three-point correlation function as 
\begin{tcolorbox}[top=-3pt, bottom=2pt]
\begin{equation}\label{eq:3Ddecompose}
    \begin{split}
       \overline{ \langle D(E_1)D(E_2)D(E_3)\rangle }&=\bar{D}(E_1)\bar{D}(E_2)\bar{D}(E_3)     +\langle D(E_1)D(E_2)D(E_3)\rangle_c \\
        &+\langle D(E_1)\rangle \langle D(E_2)D(E_3)\rangle_c+\langle D(E_2)\rangle \langle D(E_3)D(E_1)\rangle_c+\langle D(E_3)\rangle \langle D(E_2)D(E_1)\rangle_c \,.
    \end{split}
\end{equation}
\end{tcolorbox}
The fully connected three-point contribution is
\begin{equation}
    \begin{split}
        \langle D(E_1)D(E_2)D(E_3)\rangle_c&=e^{-2S_0}\delta(E_2-E_1)\delta(E_3-E_1)\langle D(E_1)\rangle       +2e^{-3S_0} K(E_1,E_2)K(E_2,E_3)K(E_3,E_1) \\
        &-e^{-3S_0}\bigg[\delta(E_2-E_3)K(E_1,E_2)^2+\delta(E_3-E_1)K(E_2,E_3)^2+\delta(E_1-E_2)K(E_3,E_1)^2\bigg] \,. 
    \end{split}
\end{equation}
It is worth mentioning here that each term in Eq.~\eqref{eq:3Ddecompose} corresponds to Euclidean wormholes with different topology (see Fig.~\ref{fig:wormholes}). We refer to these as Bra–Operator–Ket, Bra–Operator, Operator–Ket, and Bra–Ket, respectively. The disconnected contribution corresponds to the classical geometry of the AdS black hole and dominates in the classical limit $e^{S_0} \to \infty$.

For the following explicit calculations of time evolution, we will restrict to the GUE, where the CD kernel near $E_i \approx E_j$ becomes the sine kernel:
\begin{equation}
K(E_i, E_j) \approx   \frac{\sin\left(\pi (E_i-E_j)\rho(\bar{E})\right)}{\pi(E_i-E_j)} \approx K(E_j, E_i) \,.
\end{equation}
Using the limits 
\begin{equation}
\lim _{N \rightarrow \infty} \frac{\sin (N x)}{\pi x}=\delta(x), \qquad \lim _{N \rightarrow \infty} \frac{(\sin (N x))^2}{N x^2} \sim \pi \delta(x) .
\end{equation}
we can find that the CD kernel in the classical limit $N \sim e^{S_0} \to \infty$ reduces to
\begin{equation}
\lim_{e^{S_0} \to \infty}   K(E_i, E_j)   \to  \delta (E_i -E_j) \,, \qquad 
  \lim_{e^{S_0} \to \infty}   K(E_i, E_j)^2   \to  \rho(\bar{E})\delta (E_i -E_j) \,. \\
\end{equation}
Therefore, all connected parts vanish in the classical limit, \ie 
\begin{equation}
\lim_{e^{S_0} \to \infty} \langle \rho_i \rho_j \rangle_c = \lim_{e^{S_0} \to \infty} \left( \rho(\bar{E})\delta (E_i -E_j)-  K(E_i, E_j)^2   \right)  = 0  = \lim_{e^{S_0} \to \infty} \langle \rho_i \rho_j \rho_k \rangle_c  \,.  \\
\end{equation}
Thus, the classical limit of the spectral correlation functions yields
\begin{equation}
 \boxed{\textbf{Classical limit:} \lim_{e^{S_0} \to \infty} \left\langle\rho(E_i) \rho(E_j)\right\rangle =  \rho(E_i) \rho(E_j) \,,  
 \qquad \lim_{e^{S_0} \to \infty} \left\langle\rho(E_i) \rho(E_j) \rho(E_k)\right\rangle =  \rho(E_i) \rho(E_j) \rho(E_k)  \,. }
\end{equation}
From the perspective of Euclidean geometries in the gravitational path integral, this explains that the disconnected part and connected part correspond to classical and quantum contributions, respectively.

\subsection{B.2 Codimension-zero generating functions $\langle e^{-\alpha_+ \mC_+ -\alpha_- \mC_-} \rangle$}
Having obtained the three-point spectral correlation function in Eq.~\eqref{eq:3Ddecompose}, we now return to the generating function of codimension-zero holographic complexity defined in Eq.~\eqref{eq:gen0}. Applying the residue theorem to this particular pole structures, we compute each component of the generating function in correspondence with the terms in the three-point decomposition. To wit, 
\begin{equation} \label{eq:Gzerototal}
    \langle e^{-\alpha_+ \mC_+ -\alpha_- \mC_-} \rangle=G_{\rm classical}+G_{\rm bra-op}+G_{\rm op-ket}+G_{\rm bra-ket}+G_{\rm bra-op-ket} \,. 
\end{equation}
We present only the final expressions below, leaving detailed derivations to a forthcoming paper \cite{toappear}. Using the previous results for the codimension-one generating function $\langle e^{-\alpha \mC} \rangle$, we can express each component of the codimension-zero case as:
\begin{equation} \label{eq:Gzero}
\begin{split}
    G_{\rm classical}(\alpha_\pm,t)&=G_{\rm classical}(\alpha_+,t)\,G_{\rm classical}(\alpha_-,t)\\
    G_{\rm bra-op}(\alpha_\pm,t)&=G_{\rm quantum}({\alpha}_+,t)\,G_{\rm classical}(\alpha_-,t)\\
    G_{\rm op-ket}(\alpha_\pm,t)&=G_{\rm quantum}({\alpha}_-,t)\,G_{\rm classical}(\alpha_+,t)\\
    G_{\rm bra-ket}(\alpha_\pm,t)&=G_{\rm quantum}({\alpha}_++ {\alpha}_-,t)\\
    G_{\rm bra-op-ket}(\alpha_\pm,t)&=G_{\rm delta}({\alpha}_+)\,G_{\rm delta}({\alpha}_-)+G_{\rm delta}({\alpha}_+)\,G_{\rm sine}({\alpha}_-,t)+G_{\rm delta}({\alpha}_-)\,G_{\rm sine}({\alpha}_+,t)\\
    &\quad+G_{\delta_{12}K_{13}}({\alpha}_\pm,t)+G_{K_{12}K_{23}K_{31}}({\alpha}_\pm,t) \,,
\end{split}
\end{equation}
Here, $G_{\rm classical}$, $G_{\rm delta}$, $G_{\rm sine}$, and $G_{\rm quantum}$ are the same functions as in the codimension-one generating function (see Eq.~\eqref{eq:Gclassical}). The final two terms, $G_{\delta_{12}K_{13}}$ and $G_{K_{12}K_{23}K_{31}}$, arise from the fully connected {\it bra–op–ket} component and represent new contributions. It is interesting to note this decomposition precisely mirrors the one used to classify Euclidean spacetime geometries associated with codimension-one observables.

Before turning to the time evolution of codimension-zero holographic complexity, we briefly illustrate how the generating function $\langle e^{-\alpha_+ \mC_+ -\alpha_- \mC_-} \rangle$ reduces to its codimension-one counterpart. For a complete $\mC_\pm$-basis, we have
 \begin{equation}\label{eq:aminuszero}
\boxed{\lim_{\alpha_- \to 0} \langle \widehat{(e^{-\alpha_+ \mC_+}})_{\text{reg}}\cdot(\widehat{e^{-\alpha_- \mC_-}})_{\text{reg}} \rangle = \langle \text{TFD} | (\widehat{e^{-\alpha_+ \mC_+ }})_{\text{reg}} \cdot  \hat{\mathbbm{1}}  | \text{TFD} \rangle  =   \langle \text{TFD} | (\widehat{e^{-\alpha_+ \mC_+ }})_{\text{reg}} | \text{TFD} \rangle =  \langle (\widehat{e^{-\alpha \mC}})_{\text{reg}} \rangle \,.}
\end{equation}
However, in practice, we must regularize the artificial $\delta^2$-type divergence that arises from the continuum limit. As a double-check for the regularization in Eq.~\eqref{eq:twodelta}, let us consider the complete basis where the matrix element is given by Eq.~\eqref{eq:completematrixelement}. Then
\begin{equation}
\begin{split}
 \lim_{\alpha_- \to 0} \langle e^{-\alpha_+ \mC_+ -\alpha_- \mC_-} \rangle   &= \frac{e^{3S_0}}{Z}\int dE_1dE_2 dE_3 \, e^{-iE_{12}t}  \overline{\langle  D(E_1)D(E_2)D(E_3) \rangle} \times \langle E_1 | \widehat{e^{-\alpha_+ \mC_+ }}| E_3 \rangle \, \langle E_3 | \hat{\mathbbm{1}} | E_2 \rangle   \,.
\end{split}
\end{equation}
We focus on the following portion: 
\begin{equation}
e^{3S_0} \int dE_3  \, \overline{\langle  D(E_1)D(E_2)D(E_3)\rangle } \times \frac{\delta(E_2 -E_3)}{e^{S_0}D(E_2)}  = \int dE_3  \, \langle  \rho_1\rho_2 \rho_3\rangle \times \frac{\delta(E_2 -E_3)}{\rho_2}\,. 
\end{equation}
with
\begin{equation}
\left\langle\rho_i \rho_j \rho_k \right\rangle  = R_3(i,j,k)+\delta_{i,j} R_2(k,i)+\delta_{j,k} R_2(i,j)+\delta_{k,i} R_2(j,k)+\delta_{i,j} \delta_{j,k} \rho_i \,. 
\end{equation}
Many terms vanish due to the level repulsion, \ie 
\begin{equation}
 R_3(j,i,i)=0 \,, \qquad R_2(i,i) =0 \,. 
\end{equation}
Collecting the non-vanishing terms and recognizing that exactly one of the “$\delta^2$" contributions must be dropped as the trivial self-contraction, we get 
\begin{equation}
\boxed{\int dE_3  \, \overline{\langle  \rho_1\rho_2 \rho_3\rangle } \times \frac{\delta(E_2 -E_3)}{\rho_2} =  R_2\left(E_1, E_2\right)+\delta\left(E_1-E_2\right) \rho\left(E_1\right) = \overline{\left\langle\rho\left(E_1\right) \rho\left(E_2\right)\right\rangle} \,.}
\end{equation}
This immediately gives the consistent averaged result \eqref{eq:aminuszero}. Using the explicit results for $\langle e^{-\alpha_+ \mC_+ -\alpha_- \mC_-} \rangle$, we may apply the regularization in Eq.~\eqref{eq:regdelta} when taking the $\talpha_- \to 0$ limit. Similarly, other contributions involving multiple self-contractions must be regularized. 
For instance: 
\begin{equation}
   \lim_{\alpha_\pm \to 0} G^{\rm{(reg)}}_{\delta_{12}K_{13}}({\alpha}_\pm,t) \approx \left(1-\frac{T_{\mt{H}}}{3}\talpha_++\mathcal{O}(\alpha_+^2)\right)\left(1-\frac{T_{\mt{H}}}{3}\talpha_-+\mathcal{O}(\alpha_-^2)\right) \,, 
\end{equation}
and 
\begin{equation}
  \lim_{\alpha_\pm \to 0}   G^{\rm (reg)}_{K_{12}K_{23}K_{31}}({\alpha}_\pm,t) \approx 
    \begin{cases}
        2-\frac{2}{3}\left(\talpha_++\talpha_-\right)\left(T_{\mt{H}} +\frac{2t^2}{T_{\mt{H}}^2}- \frac{t^3}{T_{\mt{H}}^2}\right)\,, \qquad &t<T_{\mt{H}}\\
        2-2\left(\talpha_++\talpha_-\right)\,t\,, \qquad &t>T_{\mt{H}} \,. 
    \end{cases}
\end{equation}
As a consistency check, substituting the regularized limits
\begin{equation}
    \lim_{\alpha\rightarrow0} G^{\rm (reg)}_{\rm delta}(\alpha) = 1 \,,  \quad \lim_{\alpha_-\rightarrow0} G^{\rm (reg)}_{\delta_{12}K_{13}}(\alpha_+,\alpha_-) = -G_{\rm delta}(\alpha_+)  \,, \quad  \lim_{\alpha_-\rightarrow0} G^{\rm (reg)}_{K_{12}K_{23}K_{31}}(\alpha_+,\alpha_-) = -2G_{\rm sine}(\alpha_+)  \,, 
\end{equation}
into Eq.~\eqref{eq:Gzerototal}, we recover the expected reduction of the generating function: 
\begin{equation}\label{eq:GzerotoGone}
    \lim_{\alpha_-\to0}\langle e^{-\alpha_+ \mC_+ -\alpha_- \mC_-} \rangle=\langle e^{-\alpha_+ \mC_+} \rangle \,.
\end{equation}
Furthermore, taking $\alpha_+ \to 0$ with the same regularization likewise reproduces the unit as shown in Eq.~\eqref{eq:alphapluszero}.

\subsection{B.3 The time evolution of codimension-zero holographic complexity}
With the generating function $\langle e^{-\alpha_+ \mC_+ -\alpha_- \mC_-} \rangle$ at hand, we now derive the time evolution of codimension-zero holographic complexity by computing
\begin{equation}
  \boxed{  \langle \hat{\mC}_{\mt{Any}} \rangle=-\lim_{\alpha_\pm \to 0}(\partial_{\alpha_+}+\partial_{\alpha_-})\,\langle e^{-\alpha_+ \mC_+ -\alpha_- \mC_-} \rangle \,.}
\end{equation}
In the classical limit $e^{S_0} \to \infty$ (equivalently, $T_{\mt{H}} \to \infty$), only the disconnected contribution $G_{\rm classical}$ survives. The codimension-zero holographic complexity in this limit is given by
\begin{equation}
\begin{split}
    \langle\hat{\mC}_{\mt{Any}}\rangle|_{\rm classical}&=-\lim_{\alpha_\pm\to0}(\partial_{\alpha_+}+\partial_{\alpha_-})\,G_{\rm classical}\,(\alpha_\pm,t) \approx (M_++M_-)\,t  \,.
\end{split}
\end{equation}
This result corresponds to the long-time linear growth of holographic complexity in classical black hole geometries. Quantum corrections arise from the remaining contributions in the generating function, which correspond to Euclidean wormholes. By applying the regularization procedure as discussed previously for the $\alpha_\pm \to 0$ limit, we obtain
\begin{equation}
    \begin{split}
        \langle \hat{\mC}_{\mt{Any}} \rangle_{\rm bra-op}&=M_+\langle\hat C\rangle_{\rm quantum}\\
        \langle\hat{\mC}_{\mt{Any}}\rangle_{\rm op-ket}&=M_-\langle\hat C\rangle_{\rm quantum}\\
        \langle\hat{\mC}_{\mt{Any}}\rangle_{\rm bra-ket}&=(M_++M_-)\langle\hat C\rangle_{\rm quantum}=-
        \langle \hat{\mC}_{\mt{Any}}\rangle_{\rm bra-op-ket} \,. 
    \end{split}
\end{equation}
Notably, the quantum contributions from the {\it bra–ket} and {\it bra–operator–ket} terms exactly cancel each other.
Therefore, summing all contributions, we arrive at a factorized expression:
\begin{tcolorbox}[top=-3pt, bottom=2pt]
\begin{equation}
    \langle \hat{\mC}_{\mt{Any}} \rangle=(M_++M_-)\langle\hat{C}\rangle  
    \approx \text{Con} + 
\begin{cases}
      (M_+ +M_-) t\left(1-\frac{t}{T_{\mt{H}}}+\frac{t^2}{3 T_{\mt{H}}^2}\right)  \,,\quad  (t < T_{\mt{H}})  \\[1em]
     \frac{1}{3}(M_+ +M_-) T_{\mt{H}} \,, \quad (t \ge T_{\mt{H}}) \,,
 \end{cases} \,, 
\end{equation}
\end{tcolorbox}
where $\langle \hat{\mC}_{\mt{Any}} \rangle$ denotes the expectation value of codimension-one holographic complexity, including quantum corrections. In summary, we find that both codimension-one and codimension-zero holographic complexity exhibit the same characteristic time evolution: a long period of linear growth followed by saturation on a late-time plateau after the Heisenberg time $T_{\mt{H}}$.

\section{Appendix C: The Origin of Universal Time Evolution}\label{sec:apptime}
In this appendix, we extend our analysis beyond the Gaussian Unitary Ensemble (GUE) in random matrix theory, generalizing previous conclusions about the time evolution of holographic and quantum complexity to broader classes of quantum chaotic systems. We begin with the codimension-one observables. Using the unscaled density of states $\rho(E)$, the generating function of complexity is defined as
\begin{equation}\label{eq:spectraltwopoint02}
 \langle e^{-\alpha \mC} \rangle \equiv  \langle \text{TFD}(t)|\, \widehat{e^{-\alpha \mC }} \, |\text{TFD}(t) \rangle  = \frac{1}{Z} \int dE_idE_j \,e^{-iE_{ij}t}  \times \overline{\langle  \rho(E_i) \rho(E_j) \rangle }  \times  \langle E_i | \widehat{e^{-\alpha \mC }}| E_j \rangle  \,.
\end{equation} 
The key observation is that, by construction, all time dependence in the spectral representation arises solely from the phase factor $e^{-i E_{ij} t}$. This originates from the fact that the boundary time as a boundary condition for the state preparation determines the Lorentzian time evolution of TFD state $| \text{TFD}(t) \rangle$.

\subsection{C.1 Linear growth and pole structure}
\begin{figure}[t]
	\centering	\includegraphics[width=2.2in]{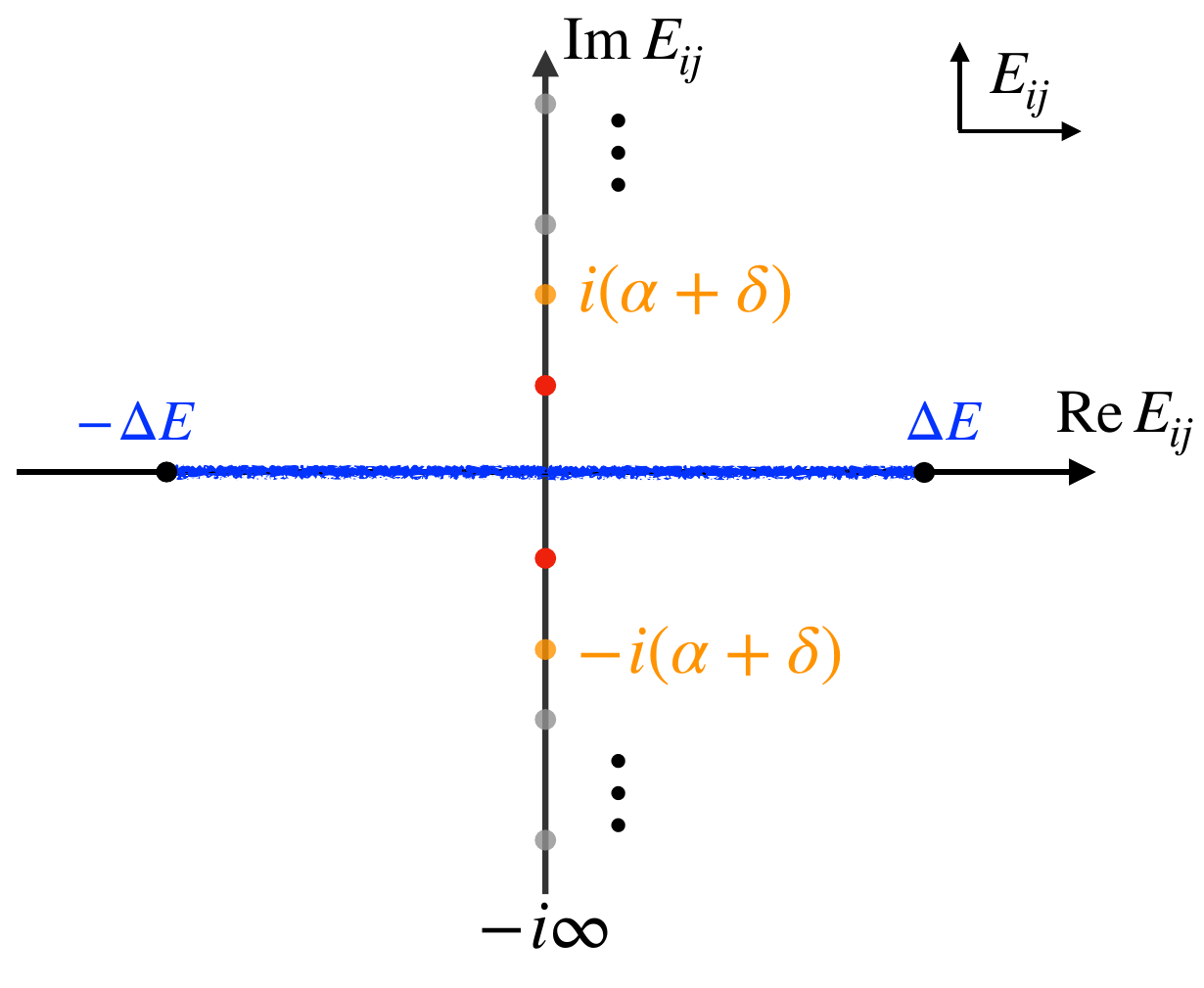}
    \includegraphics[width=2.2in]{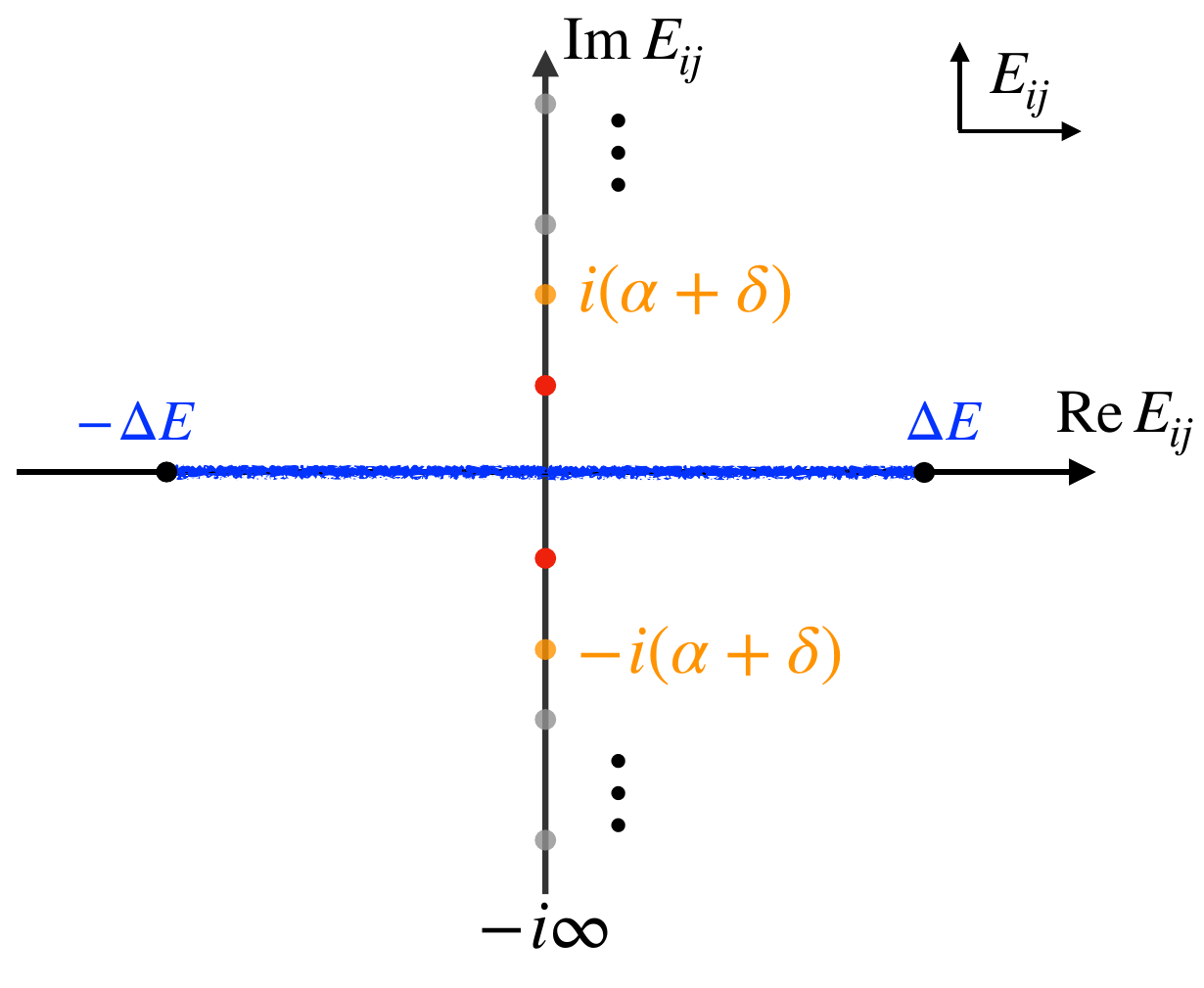}
    \includegraphics[width=2.2in]{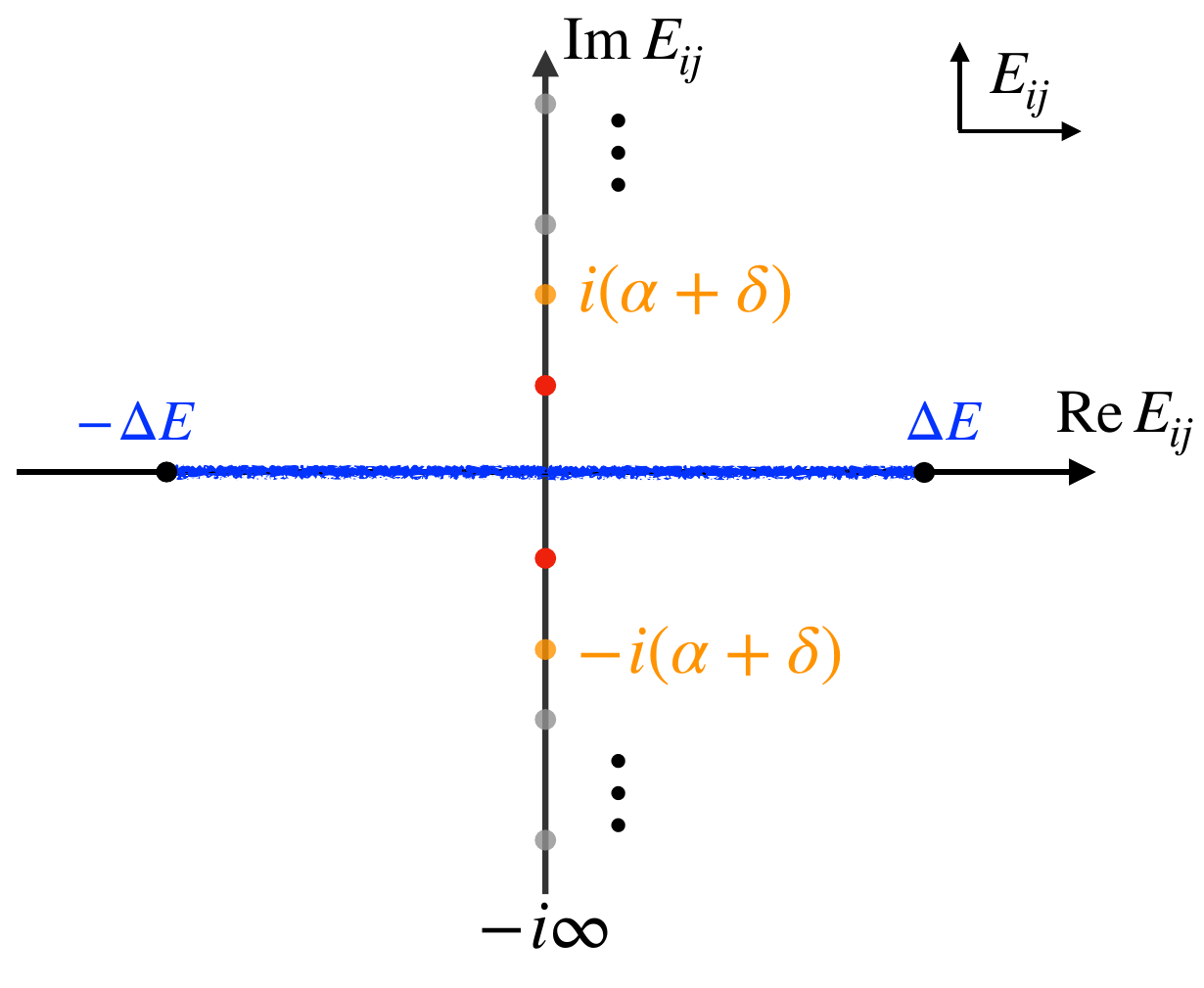}
    \caption{Analytical continuation of $\langle E_i | \widehat{e^{-\alpha \mC}} | E_j \rangle$ into the complex $E_{ij}$ plane may reveal different types of poles, each associated with a distinct time evolution of holographic complexity measures.
\textit{Left:} exponential decay; \textit{Middle:} oscillatory behavior; \textit{Right:} quadratic growth from higher-order poles.}\label{fig:otherpole}
\end{figure} 

First, let us first focus on the classical regime, where holographic complexity grows linearly up to timescales of order $t \sim e^{S_0} \to \infty$. In this limit, holographic complexity is captured entirely by the disconnected part of the spectral correlator. To probe this regime, we can take the classical limit which yields 
\begin{equation}
 \text{Classical limit:} 
 \qquad \lim_{e^{S_0} \to \infty}  \overline{ \langle  \rho(E_i) \rho(E_j) \rangle }  = \langle  \rho(E_i)  \rangle  \times  \langle  \rho(E_j) \rangle  \,. 
\end{equation}
This indicates that no quantum corrections contribute in the classical limit. Similar to previous notations, we denote the resulting contribution to the generating function by $G_{\rm classical}(\alpha, t)$. The holographic complexity derived from classical geometry is thus given by 
\begin{equation}\label{eq:Cclassical}
\boxed{ \langle \hat{C} \rangle|_{\rm classical}= \lim_{\alpha \to 0} \left(  - \partial_\alpha   G_{\rm classical}(\alpha, t)  \right) \,. } 
\end{equation}
At first glance, the time evolution might appear to depend on the detailed form of the density of states. However, assuming a smooth, non-singular $\rho(E)$, we can analytically continue $E_{ij}$ into the complex plane. The contour can be closed in the lower half-plane (Fig.~\ref{fig:contour}), where the contribution from the infinite semicircle vanishes for $t > 0$ (because of the time evolution factor $e^{-i E_{ij}t}$), and the imaginary contour $E_{ij} = \pm \Delta E + i Y$ is exponentially suppressed at large times. Applying the residue theorem, the generating function is well-approximated by the sum of the residues of the enclosed poles. This leads us to the schematic relation: 
\begin{tcolorbox}[top=-4pt, bottom=2pt]
\begin{equation}
    \text{Pole Structure} \,\Longleftrightarrow\, \text{Time Evolution}\,,
\end{equation}   
\end{tcolorbox}
as shown in the main text. In particular, we can prove that the only type of pole that leads to linear growth in Eq.~\eqref{eq:Cclassical} is a first-order pole at
\begin{equation}
 \text{first-order pole:} \qquad E_{ij} = - i M \alpha + \mathcal{O}(\alpha^2) \,. 
\end{equation}
Suppose a complex function $f(z)$ has a pole at $z = z_{\rm p}$ of order $n$. The residue of $f(z)$ around $z = z_{\rm p}$ is then 
\begin{equation}
\operatorname{Res}(f(z)) \Big|_{z=z_{\rm p}}=\frac{1}{(n-1)!} \lim _{z \rightarrow z_{\rm p}} \frac{d^{n-1}}{d z^{n-1}}\Big(\left(z-z_{\rm p}\right)^n f(z)\Big) .
\end{equation}
To proceed, let us assume that the analytically continued matrix element $\langle E_i | \widehat{e^{-\alpha \mC}} | E_j \rangle$ contains a pole at
\begin{equation}
 \text{Pole:} \qquad E_{ij} = E_{\rm p} : =  E_{\mt R} (\alpha) + i E_{\mt I}(\alpha) \,.    
\end{equation}
As a result of the residue theorem, the time evolution is completely decoded in the pole in terms of the residue of the phase factor $e^{-i E_{ij}}t$, namely 
\begin{equation}
 G_{\rm classical}(\alpha, t)  \ni \text{Res}( e^{-i E_{ij}t} ) \big|_{E_{ij} = E_{\rm p}}  \qquad \longrightarrow \qquad e^{-i E_{\mt R} t} e^{ E_{\mt I} t} \,, 
\end{equation}
where we have ignored other parts without any time dependence. In terms of the holographic complexity, its definition \eqref{eq:Cclassical} then implies 
\begin{equation}
\langle \hat{C} \rangle|_{\rm classical} \ni 
\left( i E_{\mt{R}}'(\alpha) -   E_{\mt{I}}'(\alpha) \right)  e^{-i E_{\mt R}(\alpha) t} e^{ E_{\mt I}(\alpha) t}  \bigg|_{\alpha \to 0}\,. 
\end{equation}
Importantly, this exponential behavior $e^{E_{\mt I} t}$ will dominate unless $\lim_{\alpha \to 0} E_{\mt I}(\alpha) = 0$. Similarly, the non-zero real part $E_{\mt R} (\alpha)$ would introduce oscillating time evolution. We then arrive at the necessary condition for the linear growth of holographic complexity, \ie the real and imaginary part of the pole should satisfy 
\begin{tcolorbox}[top=-3pt, bottom=0pt]
\begin{equation}
\lim_{\alpha \to 0} E_{\mt R} (\alpha) = 0  \,, \quad \text{and} \quad  \lim_{\alpha \to 0} E_{\mt I} (\alpha)  \sim - M \alpha + \mathcal{O}(\alpha^2)  \to  0 \,,
\end{equation}     
\end{tcolorbox}
where the coefficient $M$ (\ie $\talpha \approx M \alpha$ used in the main text) determines the linear growth rate. For illustrative purposes, we summarize alternative pole structures (with a finite shift $\delta$) and the corresponding time behaviors:  
\begin{equation}\label{eq:otherpoles}
G_{\rm classical}(\alpha, t) 
\approx 
\begin{cases}
 e^{-(\talpha+\delta) t} \pi   \quad \longrightarrow\quad   te^{-\delta t} \,, \qquad \text{Pole at}  \quad E_{ij} = \pm i (\talpha + \delta)   \,,\\[1em]
e^{-\talpha t + i \delta t} \pi   \quad \longrightarrow\quad   t \cos (\delta t)\,,\quad \text{Poles at} \quad E_{ij} = \pm ( - i\talpha \pm \delta )   \,. 
\end{cases}
\end{equation}
We note that poles always appear in pairs on the complex plane because the energy difference $E_{ij}$ has a $\mathbb{Z}2$ symmetry under $E{ij} \leftrightarrow -E_{ij}$ and the matrix elements are real by definition. 
Except for the first-order pole, the matrix element of the generating function may also contain some higher-order poles. However, we can find that the higher-order poles at $E_{ij}= - i \talpha$ can only yield quadratic growth rather than the expected linear growth. See the following calculation for a straightforward proof: 
\begin{equation}\label{eq:otherpoles02}
G_{\rm classical}(\alpha, t) 
\approx 
\begin{cases}
\frac{1}{2} \pi  e^{-\alpha t} \left(\frac{1}{\alpha }+t\right)  \quad \longrightarrow\quad  \frac{\pi }{2 \alpha ^2}+\frac{\pi  t^2}{4} + \cdots\,,\qquad \text{Second-order pole at} \quad E_{ij} = \pm (i\alpha )   \,,\\[1em]
\frac{\sqrt{\pi } 2^{\frac{3}{2}-n} \alpha ^{\frac{5}{2}-n} t^{n-\frac{1}{2}} K_{n-\frac{1}{2}}(t \alpha )}{\Gamma (n)} \quad \longrightarrow\quad  \frac{\sqrt{\pi } \Gamma \left(n-\frac{3}{2}\right) \left(\frac{2 (3-2 n)^2}{\alpha ^2}+(5-2 n) t^2\right)}{4 \Gamma (n)} + \cdots\,, \text{nth-order pole}  \,, 
\end{cases}
\end{equation}
In summary, we conclude that the necessary and sufficient condition for long-time linear growth of holographic complexity is the presence of a specific first-order pole in the matrix element of the generating function, \ie 
\begin{tcolorbox}[colbacktitle=CornflowerBlue!10!lightgray!60!,coltitle=black,enhanced, attach boxed title to top center={yshift=-4mm},boxed title style={colframe=black},fonttitle=\bfseries, title={Classical linear growth and Pole Structure}]
\begin{equation}\label{eq:linearpole}
\langle E_i | \widehat{e^{-\alpha \mC }}  | E_j \rangle \sim \frac{\tilde{\alpha}}{(\tilde{\alpha} + i E_{ij} )(\tilde{\alpha} - i E_{ij})} + \mathcal{O}(\alpha^2)  \,.
\end{equation} 
\end{tcolorbox}

A careful reader may notice that we have thus ignored the case of four poles located at
\begin{equation}
E_{ij} = \pm ( - i\alpha \pm \delta (\alpha) )  \,, \qquad \text{with} \qquad \lim_{\alpha \to 0}\delta (\alpha) \sim \alpha^n  \to 0\,. 
\end{equation}
This configuration also leads to linear growth, since
\begin{equation}
G_{\rm classical}(\alpha, t)  \approx  e^{- \talpha t}\cos ( \delta(\alpha) t)   \qquad\to\qquad  \langle \hat{C} \rangle|_{\rm classical}  \sim  M t \left(  \cos ( \delta(\alpha) t) +  t \delta'(\alpha) \sin ( \delta(\alpha) t) \right)  \sim M t \,. 
\end{equation}
However, this form is equivalent to Eq.~\eqref{eq:linearpole} up to subleading corrections. Indeed, summing such four poles gives
\begin{equation}
\frac{1 }{2}  \left(\frac{\talpha}{\talpha ^2+(\delta +E_{ij})^2}+\frac{\talpha}{\talpha ^2+(E_{ij}-\delta )^2}\right) \approx \frac{\talpha}{\alpha^2+(E_{ij})^2} -\frac{\talpha \left(\talpha ^2-3 (E_{ij})^2\right)}{\left(\talpha ^2+(E_{ij})^2\right)^3} \delta ^2 +  \mathcal{O}(\delta^3) \,,
\end{equation}
which is equivalent to Eq.~\eqref{eq:linearpole} up to subleading corrections.

\subsection{C.2 Late-time plateau and level repulsion}
\begin{figure}[h]
	\centering	\includegraphics[width=3in]{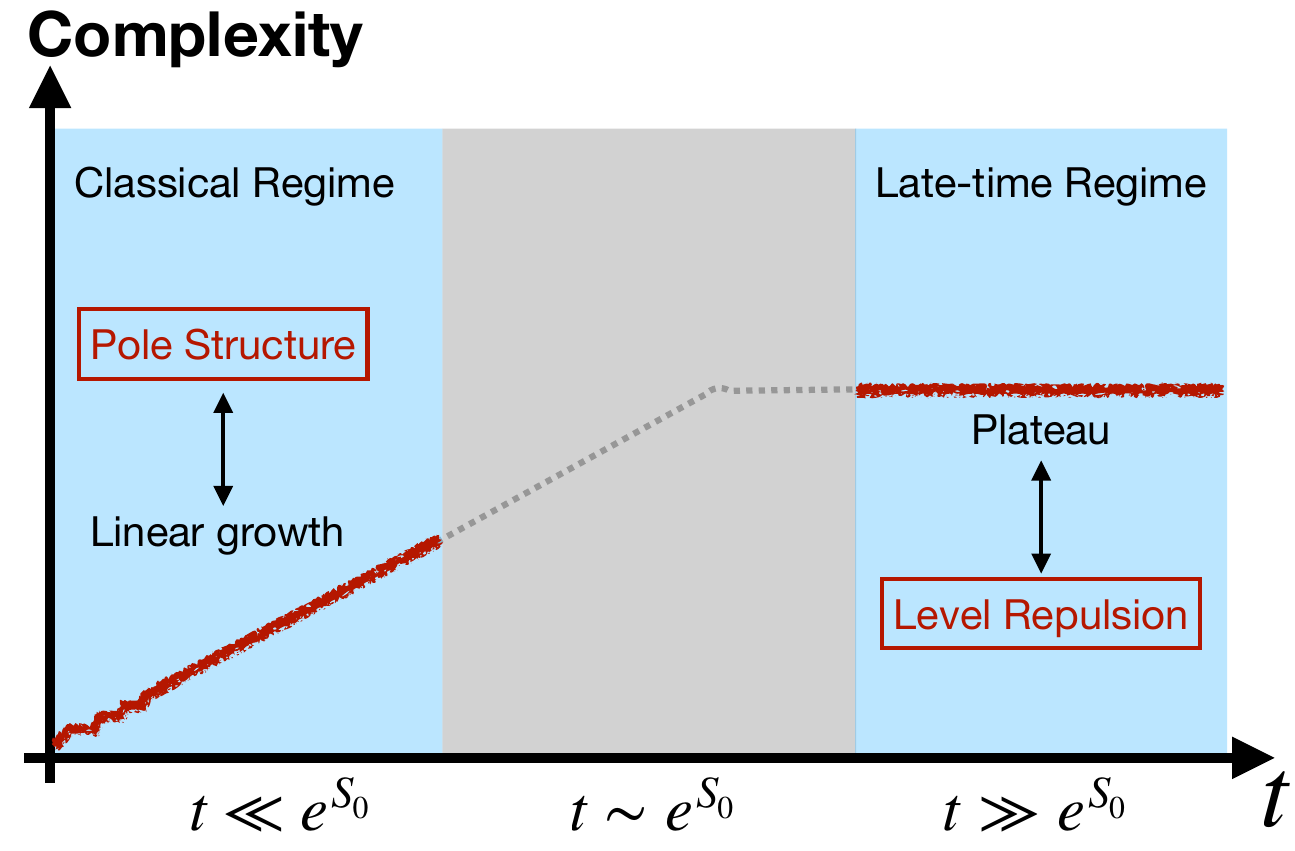}
	\caption{The universal time evolution of holographic and quantum complexity measures in chaotic systems. The linear growth and late-time plateau are governed by the particular pole structure and level repulsion.}\label{fig:universality}
\end{figure} 

To incorporate quantum corrections at finite $S_0$, we evaluate the generating function \eqref{eq:spectraltwopoint02} using the averaged spectral correlation function and the particular pole structure \eqref{eq:linearpole} established by the constraint from the linear growth of holographic complexity. Without loss of generality, we decompose the spectral correlation function into a contact term and a joint eigenvalue distribution:
\begin{equation}
\begin{split}
    \overline{ \langle \rho(E_i)\rho(E_j)\rangle }&=  R_2(E_i, E_j) +  \delta(E_i-E_j) \rho(E_i)  \,.\\
\end{split}
\end{equation}
In terms of a generic kernel $K(E_i,E_j)$, the joint eigenvalue distribution $R_2(E_i,E_j)$ can be further expressed as 
\begin{equation}
 R_2 (E_{ij}, \bar{E})  =  K\left(E_i, E_i\right) K\left(E_j, E_j\right) - K\left(E_i, E_j\right)K\left(E_j, E_i\right) = \rho_i \rho_j - (K_{ij})^2 \,.
\end{equation}
The details of the transition from linear growth to the late-time plateau depend on the precise form of the kernel $K(E_i,E_j)$, as distinct kernels can imply different choices of closed integration contours (see the sine kernel example in Fig.~\ref{fig:contour}). However, we will focus on the late-time regime and show that the late-time plateau is universal in quantum chaotic systems. It is a direct consequence of level repulsion.

The key is that the appropriate closed integration contour remains identical to that of the classical regime (the lower half-plane) after a sufficiently late time $t$, since only this choice ensures exponential decay of the time evolution factor $e^{-i E_{ij}t}$. Applying the residue theorem to the universal pole structure from Eq.~\eqref{eq:linearpole}, the late-time generating function is dominated by the residue, \ie 
\begin{equation}
\text{late times:} \qquad G_{\rm classical}(\alpha, t)  + G_{\rm quantum}(\alpha, t)  \sim  \text{Con}+  e^{-\talpha t}  \times \left(  R_2 (E_{ij}, \bar{E})\right)\big|_{E_{ij} = - i \talpha} \,.
\end{equation}
Consequently, it generates the late-time evolution of complexity measures:  
\begin{equation}
\langle \hat{C} \rangle  = \langle \hat{C} \rangle|_{\rm classical} +  \langle \hat{C} \rangle|_{\rm quantum}  \approx  M t \times  \left(  R_2 (E_{ij}, \bar{E})\right)\big|_{E_{ij} =- i \alpha = 0} \,.  
\end{equation}
Importantly, the particular pole structure at $E_{ij}=-i\tilde{\alpha}$ guarantees the equivalence of two limits: 
\begin{equation}
\boxed{ \lim_{\alpha \to 0} \overset{E_{ij} = -i\tilde{\alpha}}{=} \lim_{E_{ij} \to 0} \,. }
\end{equation}
Thus, the existence of a late-time plateau is equivalent to 
\begin{equation}
\lim_{E_{ij} \to 0 }  R_2 (E_{i}, E_j) =R_2 (E_{i}, E_i) =0 \,.   
\end{equation}
Regardless of the specific details of quantum systems, this condition is precisely a manifestation of spectral level repulsion: the probability of finding two eigenvalues arbitrarily close to each other is strongly suppressed. In summary, we conclude that the necessary and sufficient condition for the saturation of holographic complexity after the linear growth is precisely the spectral level repulsion, \viz 
\begin{tcolorbox}[colbacktitle=CornflowerBlue!10!lightgray!60!,coltitle=black,enhanced, attach boxed title to top center={yshift=-4mm},boxed title style={colframe=black},fonttitle=\bfseries, title={Late-time plateau and Level repulsion}]
\begin{equation}
   R_2 (E_i, E_i) =\lim_{E_{ij} \to 0}  \left( \rho(E_i)\rho(E_j)  
     - K(E_i, E_j) \right) =  \lim_{E_{ij} \to 0} \left(  0 +   (E_{ij})^\beta \right)=0   \,. 
\end{equation}    
\end{tcolorbox}
Here, we expanded the joint eigenvalue distribution using the Christoffel–Darboux (CD) kernel. The parameter $\beta$ is the Dyson index that characterizes the universality class of random matrices ($\beta=1,2,4$ for GOE, GUE, and GSE, respectively). All conclusions in this section straightforwardly extend to codimension-zero complexity, where three-point spectral correlation functions are involved: 
\begin{equation}
\boxed{ \overline{\left\langle\rho_i \rho_j \rho_k \right\rangle }  = R_3(i,j,k)+\delta_{i,j} R_2(k,i)+\delta_{j,k} R_2(i,j)+\delta_{k,i} R_2(j,k)+\delta_{i,j} \delta_{j,k} \rho_i \,.   }
\end{equation}
In particular, the late-time plateau remains governed by level repulsion, as it implies 
\begin{equation}
R_3(E_j,E_i,E_i)=0 = R_2 (E_i,E_i) \,.
\end{equation}
Remarkably, level repulsion is widely recognized as a hallmark of quantum chaos. This analysis thus demonstrates that the classical linear growth of holographic complexity at late times is inevitably offset by quantum corrections arising from the chaotic nature of holographic systems, as summarized in Figure \ref{fig:universality}. Translating this result into the context of AdS black hole geometries, we find that quantum chaos ensures the saturation of the size of black hole interiors, thereby establishing a profound connection between gravitational dynamics and quantum chaotic statistics. Finally, we emphasize that level repulsion, as a sufficient condition for the emergence of the late-time plateau, strictly applies only at leading orders. Subleading corrections to complexity measures could still exhibit growth. For instance, complexity might experience a logarithmic increase at late times, \eg, 
\begin{equation}
 \mC \sim   0 \times t^n + \text{Con} +   N \log(t) + \sum_{n \ge 1} \frac{N_n}{t^n} \,, 
\end{equation}
despite the fact that its time derivative asymptotically vanishes:
\begin{equation}
\lim\limits_{t \to \infty} \frac{d \mC}{dt} =\lim\limits_{t \to \infty}  \left( 0 + 0+ \frac{N}{t} - \sum_{n \ge 1} \frac{nN_n}{t^{n+1}}  \right)   = 0   \,.
\end{equation}

\end{document}